\documentclass[aps,prd,reprint,footinbib]{revtex4-2}
\usepackage{amsmath,amssymb,bm}
\usepackage{graphicx}
\usepackage{dcolumn}
\usepackage{multirow}
\usepackage{enumitem}

\usepackage[setpagesize=false,
bookmarksnumbered=true,
bookmarksopen=true,
colorlinks=false, 
pdfborder={0 0 0}  
]{hyperref}

\begin{document}

\title{Breaking the degeneracy among regular black holes with gravitational lensing}

\author{Hong Liu}
 \author{Xiaolong Liao}%
\author{Yi Zhang}%
 \email{zhangyia@cqupt.edu.cn}
\affiliation{%
 School of Electronic Science and Engineering,\\ Chongqing University of Posts and Telecommunications, Chongqing 400065, China
}%


\begin{abstract}
We examine parameter degeneracies in Culetu, Bardeen and Hayward regular black holes across lensing, shadow and quasinormal mode regimes. Our analysis reveals that while Einstein ring data yield extremely loose constraints, with the regularization parameter $q$ exceeding $\mathcal{O}(10^3)$, they fail to improve the parameter estimation when combined with strong lensing observables. In contrast, the Event Horizon Telescope observations provide remarkably tight limits: $0 \leq q < 0.0466 <0.0847$ for Culetu, $0 \leq q < 0.5115 <0.6682$ for Bardeen and $0 \leq q < 1.0258 <1.1881$ for Hayward, which shows that the strong field regime alone dominates the available parameter space. Despite these bounds, leading order geometric observables remain highly degenerate, which masks the microscopic details of non-singular cores. To break this ``macroscopic universality,'' we identify high order signatures, such as the Lyapunov exponent and subleading time delays, as sensitive probes of near horizon curvature. Crucially, we discover that the brightness hierarchy of accretion induced intensity profiles undergoes a fundamental inversion when transitioning from lensing dominated static flows to dynamics dominated infalling flows. These results demonstrate that high resolution temporal and intensity profiles are essential for distinguishing between regular black hole geometries.
\end{abstract}

\maketitle
\raggedbottom

%\tableofcontents

\section{Introduction}
Black holes (BHs) represent the most profound predictions of General Relativity (GR) \cite{Poisson:2009pwt,Misner:1973prb,Khlopov:2008qy}. Despite the remarkable success of GR theory in describing gravitational phenomena, classical black hole solutions inevitably harbor spacetime singularities where curvature invariants diverge and the classical framework ceases to be valid \cite{steinbauer2023singularity,Hawking:1973uf}. This fundamental challenge has catalyzed the development of regular black holes (RBHs) which propose a modification of the interior spacetime geometry. In regular black holes, the central singularity is replaced by a finite, non-singular core while the existence of an event horizon is preserved \cite{Bambi:2023try,Capozziello:2024ucm,Lan:2023cvz,Bonanno:2022rvo,Fan:2016hvf,culetu2013regular,bardeen1968non,Hayward:2005gi}. In most regular black hole constructions, the central region effectively mimics a de Sitter core, ensuring that curvature scalars remain finite throughout the entire manifold. Notable and influential examples are provided by the Fan-Wang black hole horizon mapping framework \cite{Fan:2016hvf}, which encompasses several well known solutions, including the Culetu \cite{culetu2013regular}, Bardeen \cite{bardeen1968non} and Hayward black holes \cite{Hayward:2005gi}. These metrics serve as robust toy models for investigating strong field gravity without the pathological issues associated with classical singularities. Consequently, regular black holes have been extensively scrutinized over the past decade across a diverse range of theoretical and observational frameworks \cite{Kurmanov:2024hpn,Isomura:2023oqf,Suliyeva:2025tsp,Karmakar:2024hng,Maeda:2021jdc,Chiba:2017nml,Ma:2025dee,Toshmatov:2018cks,Schee:2017hof,Vagnozzi:2022moj}.

Among the various observational probes of strong gravity, gravitational lensing is paramount for testing the spacetime geometry surrounding compact objects. The deflection of light by gravitational fields was a foundational prediction of GR and remains one of its most potent observational consequences \cite{Schneider:1992bmb,Weinberg1972-WEIGAC,Ghaffarnejad:2014zva}. In strong gravitational lensing, the weak deflection limit provides a fundamental analytical tool. The Einstein angle measured in this limit serves as a primary observable to constrain black hole parameters \cite{Keeton:2005jd,Gao:2024ejs}. However, when photons propagate in close proximity to a black hole, the deflection angle can diverge, which leads to the formation of relativistic images created by photons that loop around the compact object multiple times \cite{darwin1959gravity,Virbhadra:1999nm}. Bozza et al. \cite{Bozza:2001xd,Bozza:2002zj,Bozza:2008ev,Bozza:2010xqn} pioneered the strong deflection limit formalism, wherein they showed that the deflection angle diverges logarithmically as the impact parameter approaches the critical value at the photon sphere.  This analytical framework allows for the derivation of key observables, such as the angular separation and magnification ratios of relativistic images. Since its inception, the strong deflection limit has become a standard methodology for exploring modified gravity and alternative black hole candidates \cite{Gao:2021lmo,Fu:2021fxn,Tsukamoto:2016jzh,Ghosh:2022mka,Cheong:2025lwp,Vachher:2024ezs,Turakhonov:2024xfg,Jha:2023qqz,Cunha:2018acu,Manna:2018jxb,Pang:2018jpm,Zhang:2018yzr,Wang:2019cuf,Abbas:2019olp,Lu:2019ush,Islam:2020xmy,Kumar:2020sag,Chagoya:2020bqz,Ghosh:2020spb,Zhang:2021yyi,Hsieh:2021scb,Islam:2021dyk,Islam:2021ful,Boero:2021afh,Molla:2021sgw,Islam:2022ybr,Ramadhan:2023ogm,Kumar:2023jgh,Molla:2023yxn,Xie:2024dpi,Filho:2024isd,Molla:2024sig,Dong:2024alq,Zhao:2024hep,Vishvakarma:2024icz,Vachher:2024ait,Guo:2024svn,VP:2025nwe,Lobos:2025yxe,Hsieh:2021rru,Li:2025pzy,Narayan:2019imo,Kuang:2022xjp,Cardoso:2008bp}. The strong deflection regime provides a direct observational window into near horizon geometry. This approach is further bolstered by recent milestones, such as the horizon scale images of M87* and Sgr A* captured by the Event Horizon Telescope (EHT), which have inaugurated a new era for testing strong field gravity \cite{EventHorizonTelescope:2019dse,EventHorizonTelescope:2022wkp}. Furthermore, statistical analyses leveraging EHT data have demonstrated that deviation parameters in various black hole models can be constrained to narrow intervals through $\chi^2$ tests of the shadow radius (see, e.g., \cite{Kumar:2018ple,Pantig:2022ely,Afrin:2021wlj,Zhang:2023rsy}).

To further explore the internal structures and external signatures of regular black holes, this work provides a unified investigation into the gravitational lensing features of several representative models, namely the Culetu, Bardeen and Hayward black holes, within the generalized framework proposed by Fan and Wang \cite{Fan:2016hvf}. By employing Bozza's strong deflection limit formalism, we derive analytical expressions for lensing observables and characterize the resulting relativistic images and shadow structures. Furthermore, we leverage strong lensing data to perform a $\chi^2$ analysis, which allows us to impose rigorous constraints on the regularization parameter $q$. To evaluate the observational sensitivity to various regularization schemes, we also examine the shadow intensity profiles generated by both static and infalling spherical accretion flows.

Beyond static imaging, the next generation Event Horizon Telescope (ngEHT)'s enhanced resolution and multi-frequency monitoring \cite{Johnson:2023ynn,Roelofs:2022lux,Uniyal:2025uvc} are crucial for resolving higher order photon rings and accretion variations. For regular black holes, this precision breaks the ``macroscopic universality'' that masks their microscopic details. Specifically, observations at $345 GHz$ exploit a more transparent plasma regime, which bypasses the obscuration of outer layers to probe the near horizon geometry directly. These advancements provide a definitive test to distinguish non-singular spacetimes from classical GR solutions through the detection of sub-ring structures sensitive to metric corrections.

The remainder of this paper is organized as follows: Section~\ref{BH} introduces the unified framework for regular black hole geometries. Section~\ref{lensing} presents the parameter constraints derived from both the weak deflection limit (Einstein angle) and the strong deflection regime. Section~\ref{degeneracy} discusses the inherent degeneracies observed among the three regular black hole models. Section~\ref{observations} explores potential methods to break these degeneracies through multi-messenger observables. Finally, we give a conclusion in section~\ref{conclusion}.
	
%%%%%%%%%%%%%%%%%%%%%%%%%%%
\section{The regular black hole framework}\label{BH}
We consider a static, spherically symmetric spacetime with the line element
\begin{eqnarray}
	\label{ds2}
	ds^2 = -A(r)\,dt^2 + B(r)\,dr^2 + C(r)\,(d\theta^2 + \sin^2\theta\, d\phi^2),
\end{eqnarray}
where $B(r)=A(r)^{-1}$ and $C(r)=r^2$. A unified framework encompassing the Culetu, Bardeen and Hayward solutions was proposed by Fan and Wang \cite{Fan:2016hvf}.  A generalized class of Fan-Wang solutions can be written as
\begin{eqnarray}
	\label{f_unified}
	A(r) = 1 - \frac{2 M r^2}{(r^\nu + \rho^\nu)^{\mu/\nu}},
\end{eqnarray}
where $M$ is the Arnowitt-Deser-Misner (ADM) mass, $\mu$ and $\nu$ are dimensionless exponents that determine the specific model of nonlinear electrodynamics, $\rho$ denotes a characteristic length scale associated with the regularization mechanism (for instance a magnetic charge or an effective quantum gravity scale). 

Specially, the parameter $\mu$ is set to $\mu = 3$ because this specific value ensures that the metric correctly recovers the standard Schwarzschild-AdS (Anti-de Sitter) geometry in the appropriate limit while maintaining the specific asymptotic behavior required for thermodynamic consistency and power law Maxwell field integrations. 
For numerical analysis it is convenient to introduce dimensionless variables $x=r/M$ and $q=\rho/M$, which lead to
\begin{eqnarray}
	\label{A(x)}
	A(x)=1-\frac{2x^2}{(x^\nu+q^\nu)^{3/\nu}}.
\end{eqnarray}
When $q=0$, the metric reduces to the Schwarzschild black hole. The dimensionless parameter $q$ determines the scale at which the spacetime deviates from the Schwarzschild geometry.

Furthermore, different choices of the parameter $\nu$ allow this unified description to reproduce several well known regular black hole geometries where $\nu > 0$ dictates the transition between the regular inner core and the asymptotic Schwarzschild region. For the case $\nu=1$, the metric function becomes
\begin{eqnarray}
	A(x)=1-\frac{2x^2}{(x+q)^3},
\end{eqnarray}
which is algebraically equivalent to the Culetu regular black hole \cite{culetu2013regular}%
\footnote{The case $\nu = 1$ is algebraically identical to the Culetu black hole \cite{culetu2013regular}. In Refs.~\cite{Lutfuoglu:2026fks,Konoplya:2025ect}, this form is referred to as `` black hole solutions free of curvature singularities, sourced by dark matter halos described by galactic density profiles.'' For simplicity, we denote the $\nu = 1$ case as the ``Culetu black hole.''}. In this scenario, the dimensionless parameter $q$ is interpreted as a length scale characterizing the modification of the gravitational field near the center and smoothing the curvature singularity that would otherwise appear in the Schwarzschild solution. Consequently, the spacetime geometry interpolates smoothly between a regular inner core and the standard vacuum exterior.
When $\nu=2$, the metric corresponds to the Bardeen black hole \cite{bardeen1968non}, which reduces to
\begin{eqnarray}
	A(x)=1-\frac{2x^2}{(x^2+q^2)^{3/2}}.
\end{eqnarray}
The Bardeen solution is originally proposed as the first example of a regular black hole which can be interpreted as a self-gravitating magnetic monopole arising from nonlinear electrodynamics. In this context, the parameter $q$ is associated with the magnetic charge of the configuration. 
Finally, for $\nu=3$, the metric Eq. (\ref{A(x)}) reproduces the Hayward black hole \cite{Hayward:2005gi} which is
\begin{eqnarray}
	A(x)=1-\frac{2x^2}{x^3+q^3}.
\end{eqnarray}
This model is frequently regarded as an effective description of quantum gravity corrections near the black hole center. Here, the parameter $q$ sets the size of the de Sitter-like core that replaces the classical singularity, providing a minimal length scale at which gravitational collapse is halted.

Although these models differ in physical interpretation and internal structure, they share several key features. 
Firstly, the parameter $q$ represents a characteristic regularization scale controlling deviations from the Schwarzschild geometry. Physical consistency requires $q \ge 0$, as negative values may introduce additional singularities in the metric function, which contradicts the purpose of regular black hole constructions. 
Secondly, all three geometries reduce to the Schwarzschild solution in the limit $q \to 0$, which indicates that they can be regarded as regular deformations of Schwarzschild spacetime, with $q$ quantifying the strength of corrections. 
Finally, despite distinct near origin behaviors, the Culetu, Bardeen and Hayward metrics share the same asymptotic structure. At large radii ($r \gg q$), the metric function approaches $A(x)\approx 1-2/x+\mathcal{O}(x^{-2})$, so that the asymptotically flat Schwarzschild geometry is recovered.
%%%%%%%%%%%
%%%%%%%%%%%%%%%
\begin{table*}
		\caption{
			Marginalized $1\sigma$ and $2\sigma$ confidence level constraints on parameters of Culetu, Bardeen and Hayward black holes. The constraints are derived from weak lensing observations of ESO325-G004, strong lensing observations of Sgr A* and M87* (shadow angular radius $\theta_d$) and a combined analysis of both weak and strong lensing data.
		}
		\begin{ruledtabular}
		\begin{tabular}{ccccc}
			BH types & Parameters & Einstein ring constraints & Strong lensing constraints & Joint lensing constraints\\
			\hline
			\multirow{2}{*}{Culetu BH}
			& $q_{1\sigma}$ & $0.7978\times10^3$ & $0.0466$ & $0.0466$\\
			& $q_{2\sigma}$ & $1.0413\times10^3$ & $0.0847$ & $0.0847$\\
			\multirow{2}{*}{Bardeen BH}
			& $q_{1\sigma}$ & $5.4340\times10^4$ & $0.5115$ & $0.5115$\\
			& $q_{2\sigma}$ & $7.9982\times10^4$ & $0.6682$ & $0.6682$\\
			\multirow{2}{*}{Hayward BH}
			& $q_{1\sigma}$ & $1.3560\times10^5$ & $1.0258$ & $1.0258$\\
			& $q_{2\sigma}$ & $1.7546\times10^5$ & $1.1881$ & $1.1881$\\					
		\end{tabular}
	\end{ruledtabular}
		\label{sigmas}
\end{table*}
%%%%%%%%%%%%%%%%%
%%%%%%%%%%%%%%
\begin{figure*}[!htb]
	\includegraphics[width=8cm]{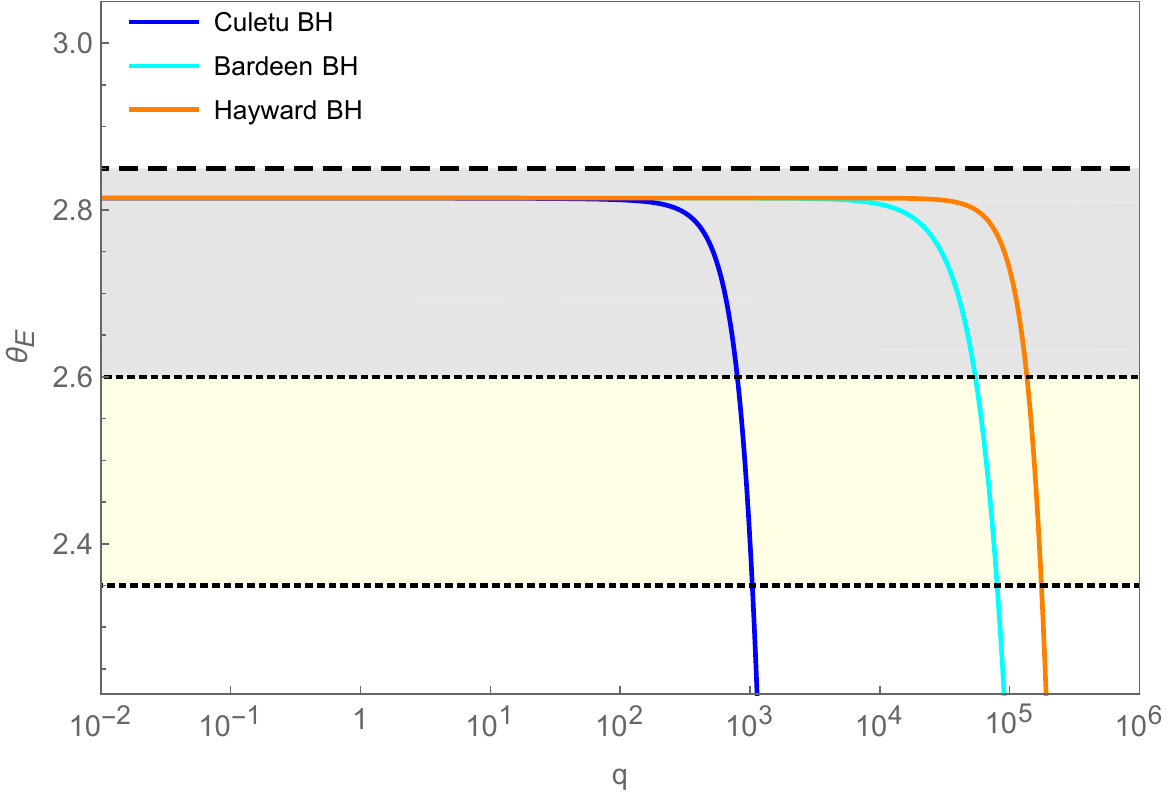}
	\includegraphics[width=8cm]{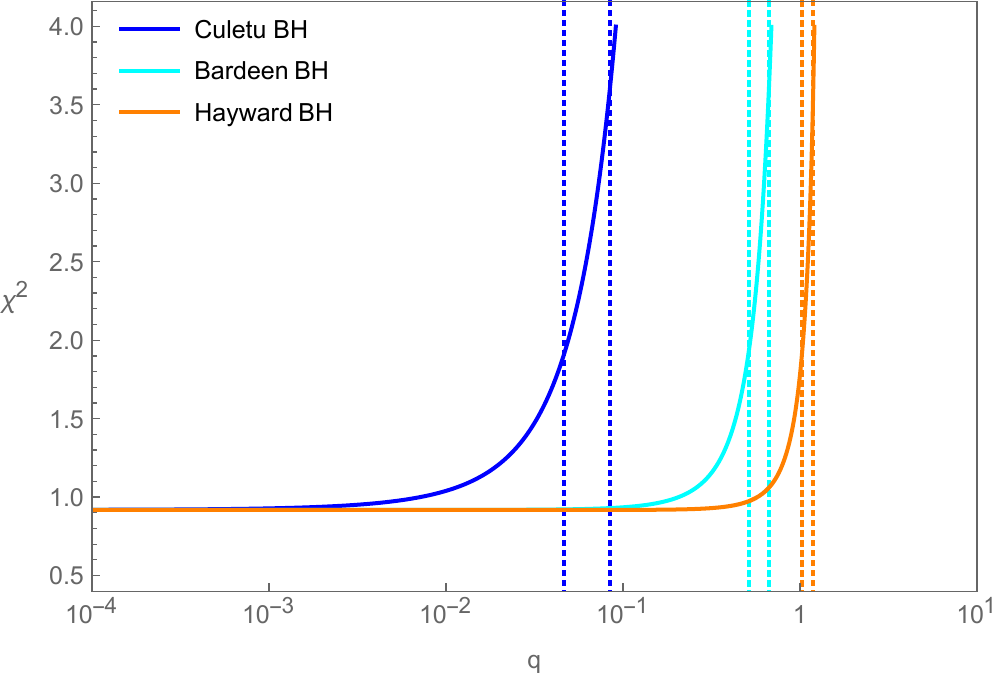}
	\caption{The angular radius of Einstein ring $\theta_E$ as functions of parameter $q$ and the $\chi^2$ test results for Culetu, Bardeen and Hayward black holes. In left one, the data from galaxy ESO325-G004 are $\theta_E^{\rm obs}=2.85^{+0.55}_{-0.25} as$ \cite{Smith:2005pq,Gao:2024ejs}. The black dashed line denotes best value. The dotted lines indicate $1\sigma$ and $2\sigma$ ($\theta_E=2.35 as$) confidence levels, while the gray and yellow shaded regions represent corresponding uncertainties in $\theta_E$. To provide a clearer visualization of constrained regions, we have restricted the display range of $\theta_E$ by excluding its upper values. In the right one, the constraints are derived from strong gravitational lensing data. The observed shadow angular radius for M87* and Sgr A* are $21 \pm 1.5\,\mu\text{as}$ and $24.35 \pm 3.5\,\mu\text{as}$, respectively. The $x$-axis utilizes a logarithmic scale to capture high resolution details of the deviation parameter $q$. Given that the Schwarzschild limit ($q = 0$) is the best fit result, we employ a practical lower limit of $10^{-4}$ to circumvent the singularity of the logarithmic axis at zero. This serves as a numerical proxy to represent the regime where $q \geq 0$, allowing for a continuous visualization from near Schwarzschild conditions to more extreme deviations.}
	\label{chi2}        
\end{figure*}
%%%%%%%%%%%%%%%%%%%
%%%%%%%%%%%%%%%
\section{Gravitational lensing by regular black holes}\label{lensing}
We investigate the gravitational lensing signatures of Culetu, Bardeen and Hayward regular black holes in both the weak and strong field regimes. We consider photon motion confined to the equatorial plane ($\theta = \pi/2$) of a static, spherically symmetric spacetime. For a photon with a distance of closest approach $x_0$, the total deflection angle is given by \cite{Bozza:2002zj}
\begin{eqnarray}
	\alpha(x_0) = - \pi + \tilde{I}(x_0),
\end{eqnarray}
where
\begin{eqnarray}
	\tilde{I}(x_0) = 2 \int_{x_0}^{\infty} \frac{x_0}{x} \frac{dx}{\sqrt{x^2 A(x_0) - x_0^2 A(x)}}.
\end{eqnarray}
%%%%%%%%%%%%%%%%%%%%%%%%

%%%%%%%%%%%%%%%%%%%%%%%
\subsection{Constraints from Einstein ring}
In the weak field limit ($x_0 \gg 1$), for a system where the source, lens and observer are nearly perfectly aligned, the Einstein ring radius $\theta_E$ is defined as \cite{Bozza:2008ev}
\begin{eqnarray}
	\theta_E = \frac{D_{LS}}{D_S}\alpha(b),
\end{eqnarray}
where the impact parameter is given by $b = D_L \theta_E$. We derive the analytical expressions for the Einstein ring in \ref{appweak} and utilize the observed Einstein radius of the galaxy ESO325-G004, $\theta_E^{\rm obs} = 2.85^{+0.55}_{-0.25}\,\text{as}$ \cite{Smith:2005pq}, to constrain the regularization parameter $q$ for the Culetu, Bardeen and Hayward black hole models.

The comparison between theoretical predictions and observational data is presented in the left panel of Fig.~\ref{chi2}. As illustrated, $\theta_E$ decreases monotonically as $q$ increases across all three models. The specific constraints for the $1\sigma$ and $2\sigma$ confidence levels are summarized in Table~\ref{sigmas}. Notably, the resulting bounds on $q$ for the Culetu, Bardeen and Hayward black holes are of the order $\mathcal{O}(10^{3})$, $\mathcal{O}(10^{4})$ and $\mathcal{O}(10^{5})$, respectively.
Einstein ring constraints are loose ($q \gg 1$) because the parameter $q$ provides only a negligible, rapidly decaying correction to the mass dominated deflection at galactic scales.
%%%%%%%%%%%%%%%%%%%%%%%%%%%%

%%%%%%%%%%%%%%%%%%%%%%%%%%
\subsection{Constraints from strong gravitational lensing}
In the strong field regime, particularly near the photon sphere, we adopt the analytical formalism developed by Refs. \cite{Bozza:2001xd,Bozza:2002zj,Bozza:2008ev,Bozza:2010xqn} which is briefly presented in \ref{appBozza}. The appearance of the black hole shadow is governed by the behavior of null geodesics in this strong field region. We first examine the theoretical shadow radii ($b_m$ and $R_{sh}$) and their corresponding angular observables ($\theta_\infty$ and $\theta_d$) to establish a comparative baseline. This approach allows us to quantify how variations in the internal structure manifest in the observable shadow profile.

%%%%%%%%%%%%%%%%%%%%%%%%
\subsubsection{The photon sphere radius $x_m$, critical impact parameter $b_m$ and shadow radius $R_{sh}$}
The shadow boundary of a static, spherically symmetric black hole is determined by unstable photon orbits. The photon sphere radius $x_m$ is the largest root of \cite{Bozza:2002zj}
\begin{eqnarray}
	\frac{C'(x)}{C(x)} =\frac{A'(x)}{A(x)}.
\end{eqnarray}
The critical impact parameter $b_m$, which corresponds to the photon sphere and separates captured trajectories from those that escape to infinity, is given by
\begin{eqnarray}
	\label{bm}b_m = \sqrt{\frac{C(x_m)}{A(x_m)}}.
\end{eqnarray}
For a distant observer, the geometric shadow radius $R_{sh}$ in the celestial plane $(X, Y)$ is numerically identical to the critical impact parameter \cite{cunningham1972}
\begin{eqnarray}
	\label{castEq}X^2 + Y^2 = R_{sh}^2.
\end{eqnarray}
Because the shadow boundary is formed by photons asymptotically approaching the unstable orbit at $b=b_m$, then $R_{sh} = b_m$. And these regular black holes are spherically symmetric, the shadow is a perfect disk with an area
\begin{eqnarray}
	\tilde{A} = \pi R_{sh}^2.
\end{eqnarray}
%%%%%%%%%%%%%%%%%%%%%

%%%%%%%%%%%%%%%%%%%%%%%%
\subsubsection{The asymptotic angular position $\theta_\infty$ and shadow angular radius $\theta_d$}
To connect our theoretical results with EHT observations, we present the angular size of black hole shadow. The asymptotic angular position of relativistic images is determined by the critical impact parameter
\begin{eqnarray}
	\theta_\infty = \frac{b_m}{D_L}.
\end{eqnarray}
The angular radius of shadow, as typically reported by the EHT, is given by \cite{Abdujabbarov:2016hnw}
\begin{eqnarray}
	\theta_d = \frac{1}{D_L}\sqrt{\frac{\tilde{A}}{\pi}}.
\end{eqnarray}
For static and spherically symmetric (regular) black holes, the shadow boundary is fully determined by the photon sphere and the shadow forms a perfect circular disk. For an observer located at infinity, the celestial coordinate satisfies $\sqrt{X^2+Y^2}=R_{sh} = b_m$, leading to $\theta_\infty = \theta_d$.
This equality highlights that, in the asymptotic limit, both observables are governed by the same geometric scale set by the unstable photon orbit. In our analysis, $\theta_d$ is adopted as the primary observational constraint on the model parameter $q$.
%%%%%%%%%%%%%%%%%%%%%%%%%%%
%%%%%%%%%%%%%%%
\begin{figure*}[!htb]
	\includegraphics[width=8cm]{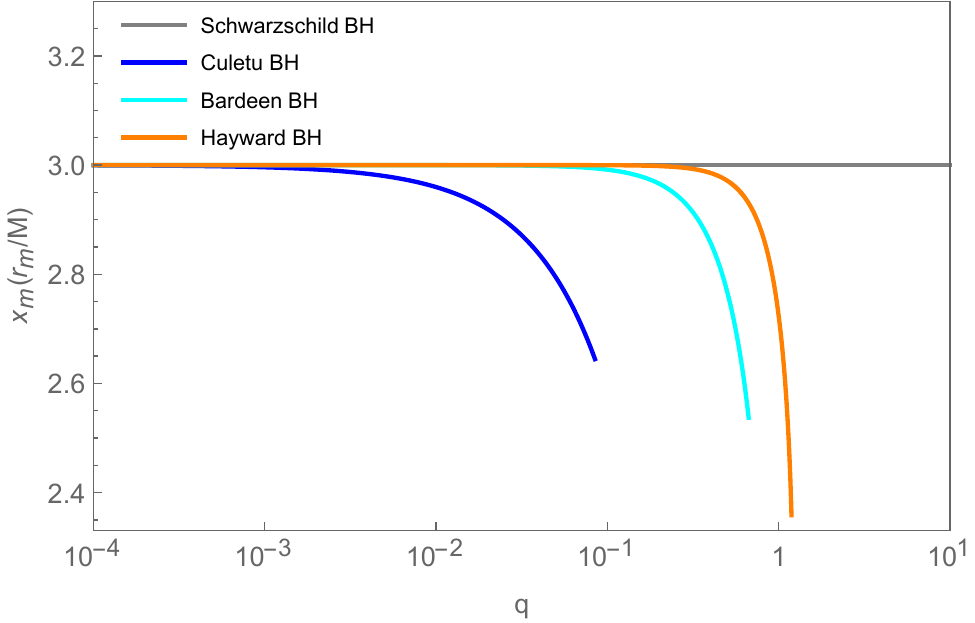}
	\includegraphics[width=8cm]{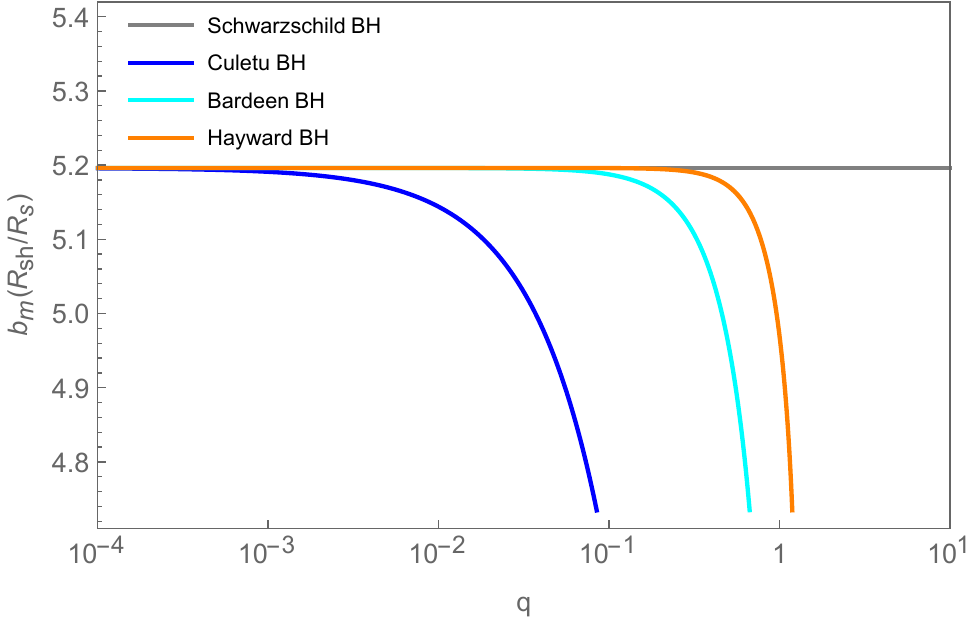}
	\caption{Evolution of photon sphere radius $x_m(r_m/M)$ and critical impact parameter $b_m(R_{sh}/R_s)$ as a function of parameter $q$. Following the strong lensing constraints established in Table \ref{sigmas}, the region for $q$ is defined as $10^{-4} < q < q_{2\sigma}$ with the lower limit effectively capturing the $q=0$ case.}
	\label{rmb}        
\end{figure*}
%%%%%%%%%%%%%%%	
%%%%%%%%%%%%%%%
\begin{table*}
		\caption{Numerical values of shadow radius $R_{sh}$ (impact parameter $b_m$), angular observables $\theta_{d}$ ($\theta_{\infty}$), time delay leading term $\Delta  T^0_{2,1}/\Delta  T^0_{3,2}/\Delta  T^0_{4,3}$ and QNM frequency $\Omega_m$. Constraints are presented at $1\sigma$ and $2\sigma$ confidence levels, denoted by $q_{1\sigma}$ and $q_{2\sigma}$. $R_{sh}$ and $b_m$ are expressed in units of gravitational radius $R_s = GM_{\bullet}/c^2$ where $M_{\bullet}$ is the mass of black hole. The $\theta_{d}$ and $\theta_{\infty}$ are given in units of micro-arcseconds ($\mu$as).}
		\begin{ruledtabular}
		\begin{tabular}{ccccc}
			\multirow{2}{*}{Parameters}&  \multirow{2}{*}{BH   type}&\multirow{2}{*}{Schwarzschild BH}&\multicolumn{2}{c}{Culetu/Bardeen/Hayward BH}\\
			&&&$ q_{1\sigma}  $&$ q_{2\sigma}$\\
			\hline
			$R_{sh}/R_s(b_m/R_s)$&$-$ &$ 5.196 $ &$ 4.948 $&$ 4.735 $\\
			\multirow{ 2}{*}{$\theta_d/ \mu as(\theta_\infty /\mu as)$}&Sgr A* &$ 26.71 $& $ 25.44  $&$ 24.34 $\\
			& M87* &$ 19.91 $ &$ 18.96  $&$ 18.14 $\\
			$\Delta  T^0_{2,1}/\Delta  T^0_{3,2}/\Delta  T^0_{4,3}$&$-$&$ 32.65 $&$ 31.09 $&  $ 29.75 $\\
			$\Omega_m$&$-$&$ 0.1925 $&$ 0.2021 $&  $ 0.2112 $\\
			
		\end{tabular}
	\end{ruledtabular}
		\label{tabtheta}
\end{table*}
%%%%%%%%%%%%%%%%%
%%%%%%%%%%%%%%%%%%%
\begin{figure*}[!htb]
	\includegraphics[width=8cm]{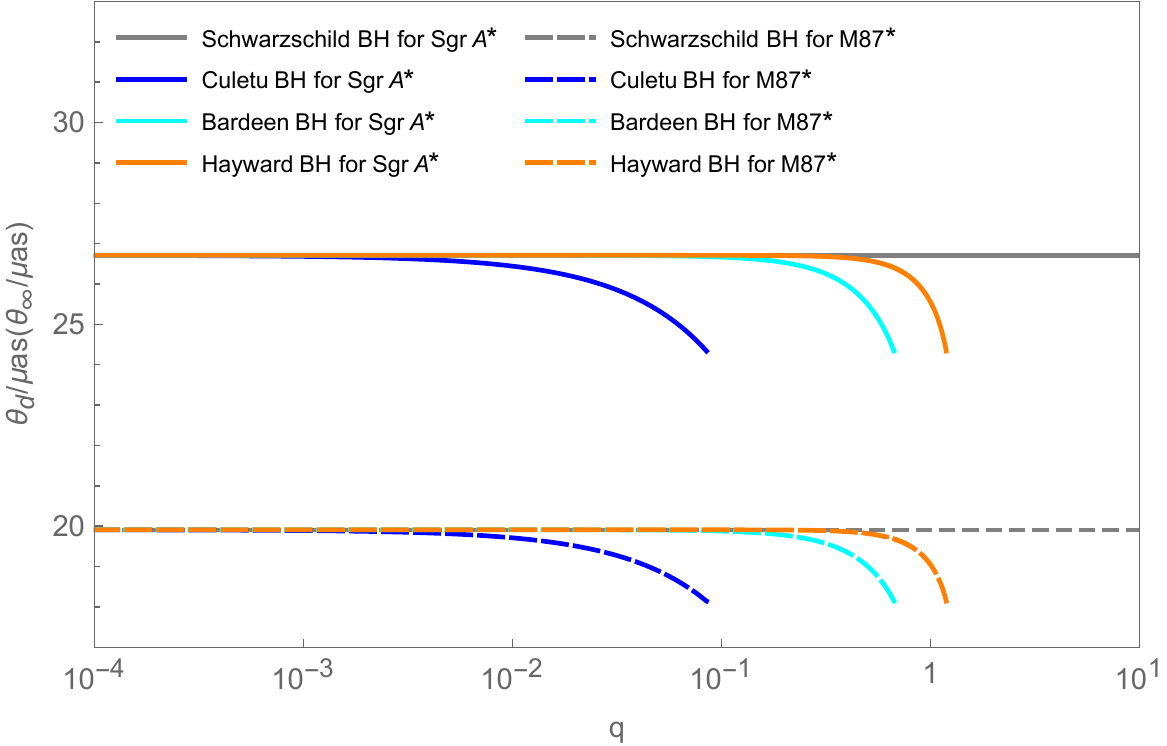}
	\includegraphics[width=8cm]{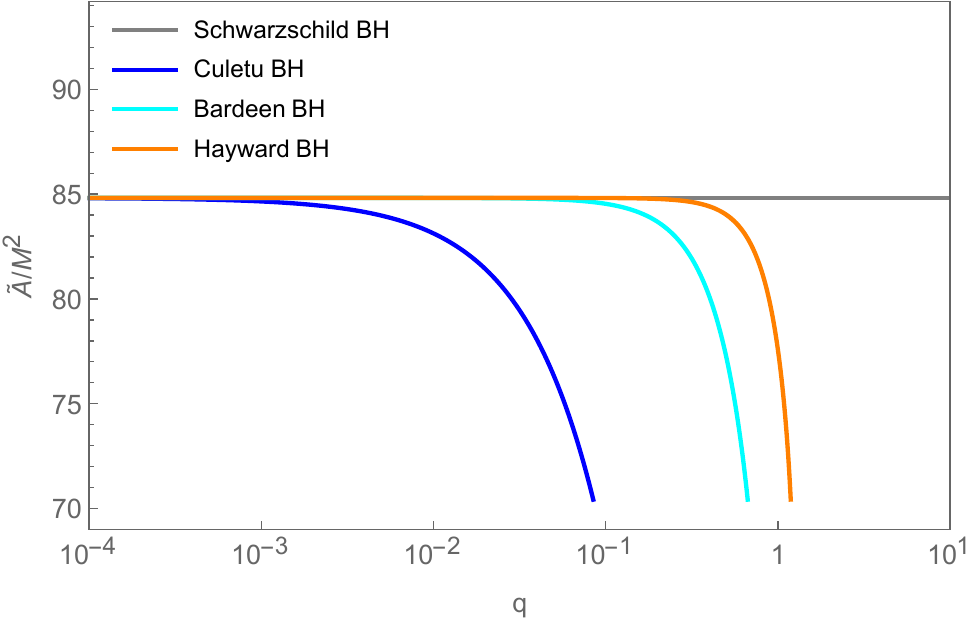}
	\includegraphics[width=8cm]{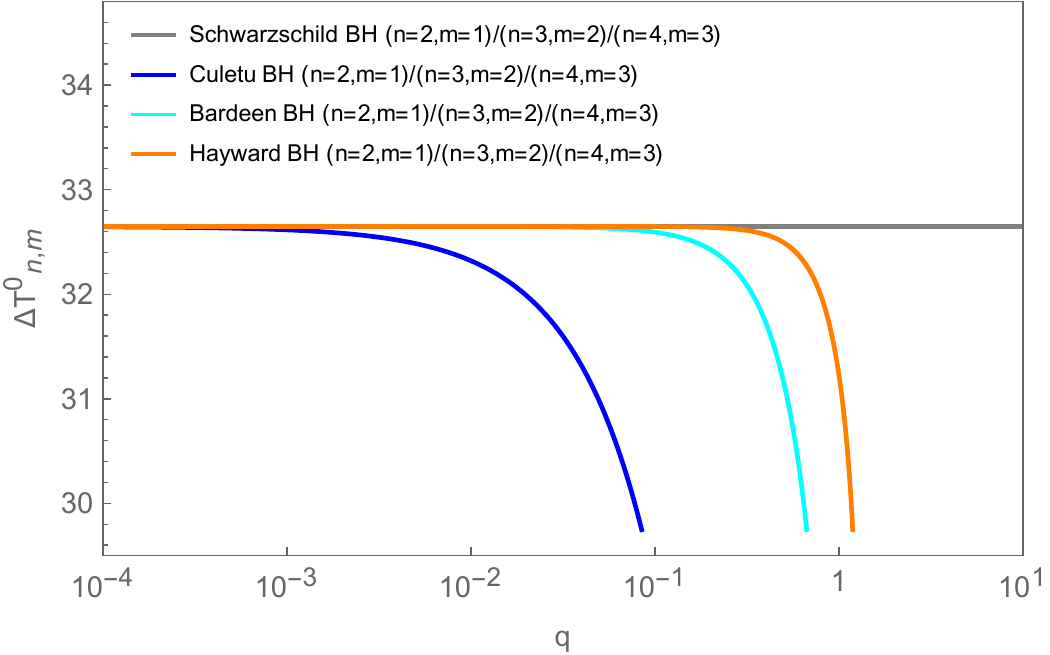}
	\includegraphics[width=8cm]{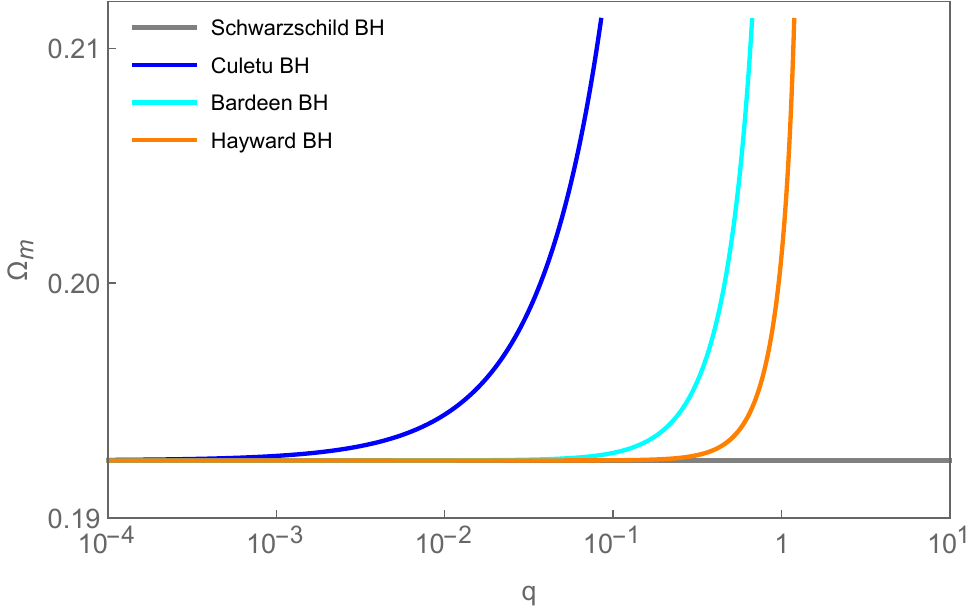}\\
	\caption{Same as Fig. \ref{rmb}, but for the angular observable $\theta_{d}/\mu as(\theta_{\infty}/\mu as)$, shadow area $\tilde{A}/M^2$, time delay leading order term $\Delta T^0_{n,m}$ and QNM frequency $\Omega_m$. }
	\label{theta}
\end{figure*}
%%%%%%%%%%%%%%%%%%%

%%%%%%%%%%%%
\begin{figure*}[!htb]
	\includegraphics[width=8cm]{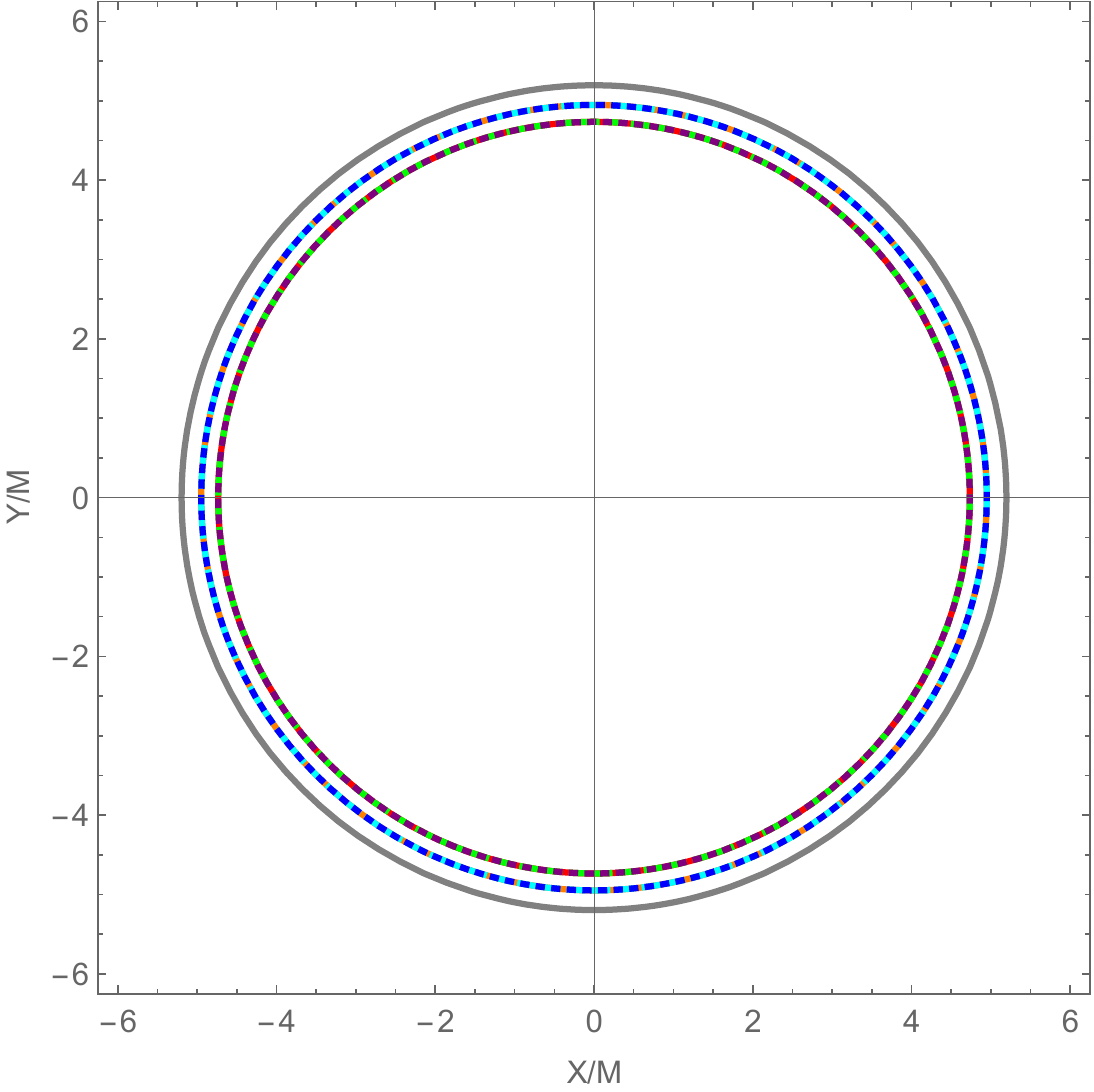}
	\caption{Shadow characteristics for the Culetu/Bardeen/Hayward black hole models. The shadows cast in the celestial plane $(X/M, Y/M)$ are illustrated for the $1\sigma$ and $2\sigma$ constraints on the deviation parameter $q$. The outermost boundary (gray line) represents the standard Schwarzschild shadow ($q=0$). The intermediate boundaries correspond to the $1\sigma$ constraints, where the shadows for the Culetu (blue), Bardeen (cyan) and Hayward (orange) models are effectively degenerate and overlap. Similarly, the innermost boundaries represent the $2\sigma$ constraints for the Culetu (purple), Bardeen (green) and Hayward (red) models which also show significant overlap. }
	\label{cast}
\end{figure*}
%%%%%%%%%%%%
%%%%%%%%%%%%%%%%%
\begin{table*}
		\caption{Numerical values of strong deflection lensing parameters ($x_m, \bar{a}, \bar{b}$), observables ($s, r$), time delay parameters ($\Delta T_{n,m}, \Delta T^1_{n,m}$) and QNM Lyapunov exponent $\lambda$. The $s$ is given in nano-arcseconds ($nas$), $r$ is in magnitudes ($mag$). Time delays are expressed in units of $GM_{\bullet}/c^{3}$, where $M_{\bullet}$ is the mass of black hole (such as for Sgr A*, $GM_{\bullet}/c^3 \approx 21.23$~s).}
		\begin{ruledtabular}
		\begin{tabular}{ccccccccc}
			\multirow{2}{*}{Parameters}&  \multirow{2}{*}{BH   type}&\multirow{2}{*}{Schwarzschild BH}&\multicolumn{2}{c}{Culetu BH}&\multicolumn{2}{c}{Bardeen BH}&\multicolumn{2}{c}{Hayward BH}\\
			&&&$ q_{1\sigma}   $&$ q_{2\sigma} $&$ q_{1\sigma}   $&$ q_{2\sigma} $&$ q_{1\sigma} $&$ q_{2\sigma}   $\\
			\hline
			$x_m(r_m/M)$&$-$&$ 3.000 $ & $ 2.809 $&$ 2.645 $& $ 2.757 $ &$ 2.537 $  & $ 2.694 $ &$ 2.358 $\\
			$\bar{a}$ &$-$ &$ 1.000 $&$ 1.034 $&$ 1.069 $ &$ 1.095 $&$ 1.217 $ &$ 1.207 $&$ 1.768 $\\
			$\bar{b} $&$-$ &$ -0.4002 $&$ -0.3942 $&$ -0.3904 $&$ -0.4616 $&$ -0.5840 $ &$ -0.7010 $&$ -2.357 $\\
			\multirow{ 2}{*}{$s/nas $}  &Sgr A*  &$ 33.43 $ &$ 39.97 $&$ 47.19 $&$ 53.82 $  &$ 80.09 $&$ 78.12 $  &$ 183.8 $\\
			&  M87*  &$ 24.92 $&$ 29.80  $  & $ 35.18 $&$ 40.09  $& $ 64.12 $ &$ 58.23  $  & $ 137.0 $\\
			$r_{mag}/mag$&$-$ &$ 6.822 $& $ 6.595 $&$ 6.384 $ & $ 6.229 $   &$ 5.607 $& $ 5.651 $   &$ 3.858 $\\
			$\Delta  T^1_{2,1}(10^{-3})  $ &$-$ &$ 497.5 $& $ 546.3 $&  $ 596.8 $& $ 665.1 $ &  $ 895.7 $& $ 866.9 $ &  $ 1709 $\\
			$\Delta T^1_{3,2}(10^{-4}) $&$-$ &$ 215.0 $& $ 262.0 $&$ 315.4 $& $ 377.6 $ &$ 677.2 $& $ 642.3 $ &$ 2892 $\\
			$\Delta T^1_{4,3}(10^{-5}) $&$-$ &$ 92.90  $&$ 125.7 $&  $ 166.7 $&$ 214.6 $&  $ 512.0 $&$ 475.9 $&  $ 4893 $\\
			$\lambda$&$-$&$ 0.0588 $&$ 0.0607 $& $ 0.0623 $&$ 0.0588 $   & $ 0.0579 $&$ 0.0557 $   & $ 0.0454 $\\
			
		\end{tabular}
		\end{ruledtabular}
		\label{tabqnm}
\end{table*}
%%%%%%%%%%%%%%%%%
%%%%%%%%%%%%%%%
\begin{figure*}[!htb]
	\includegraphics[width=8cm]{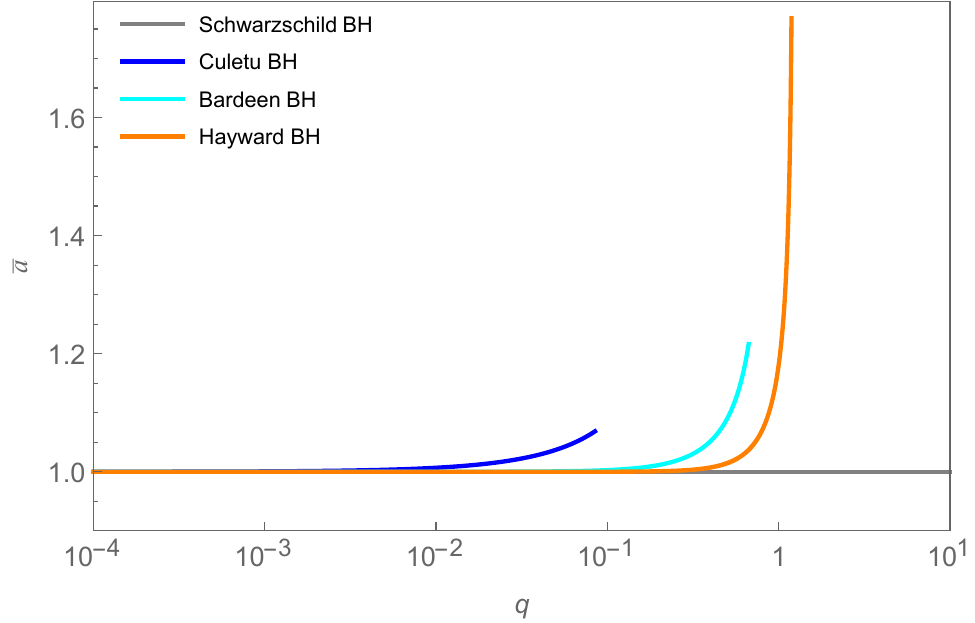}
	\includegraphics[width=8cm]{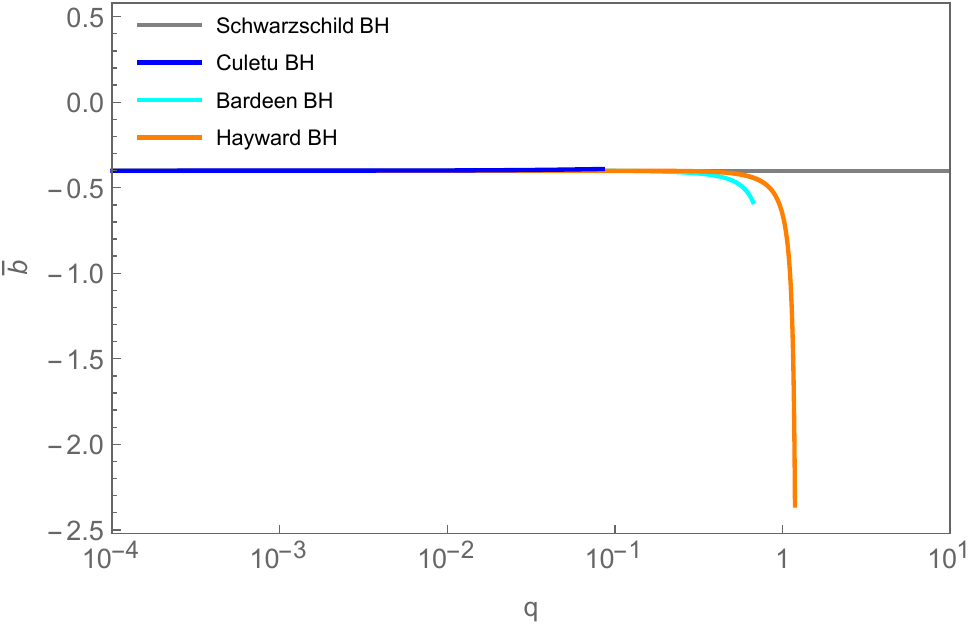}\\
	\caption{Same as Fig. \ref{rmb}, but for the lensing coefficients $\bar{a}$ and $\bar{b}$.}
	\label{a}        
\end{figure*}
%%%%%%%%%%%%%%%	
%%%%%%%%%%%%%%%%%%
\begin{figure*}[!htb]
	\includegraphics[width=8cm]{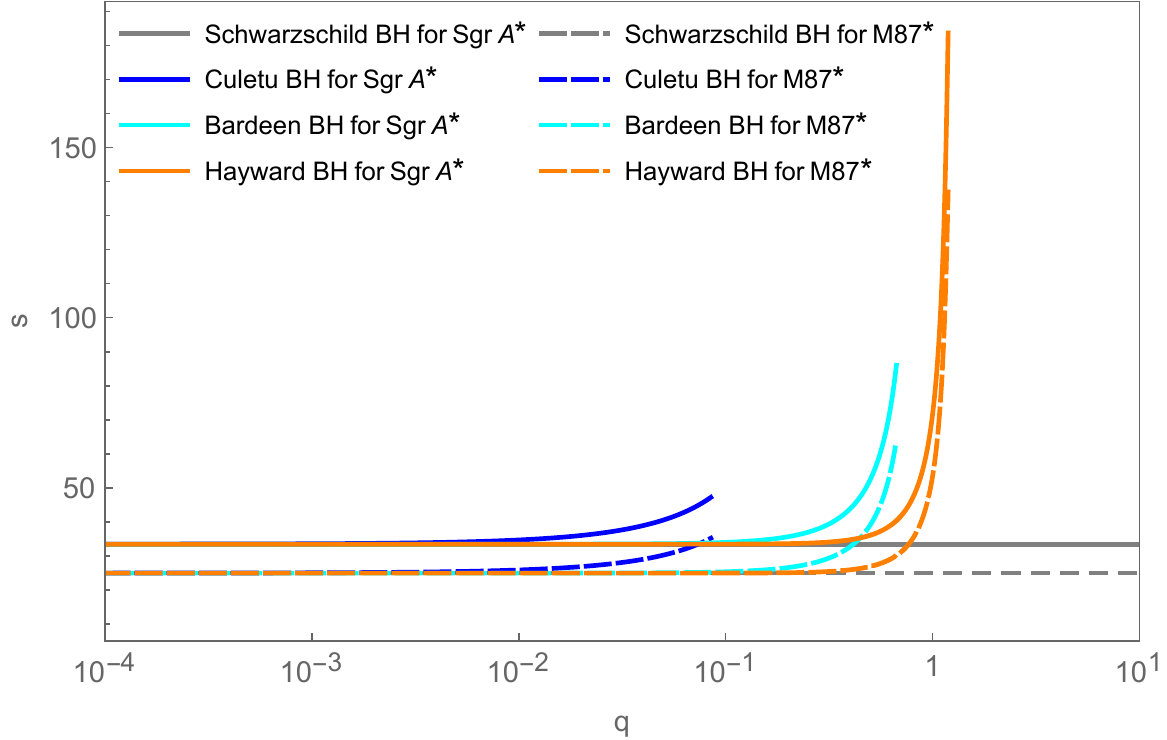}
	\includegraphics[width=8cm]{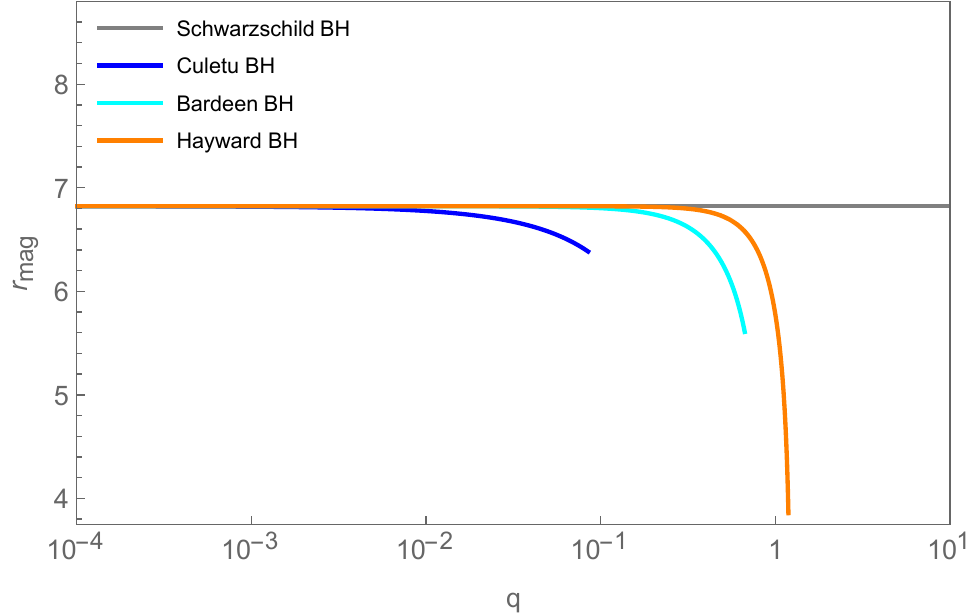}\\
	\includegraphics[width=8cm]{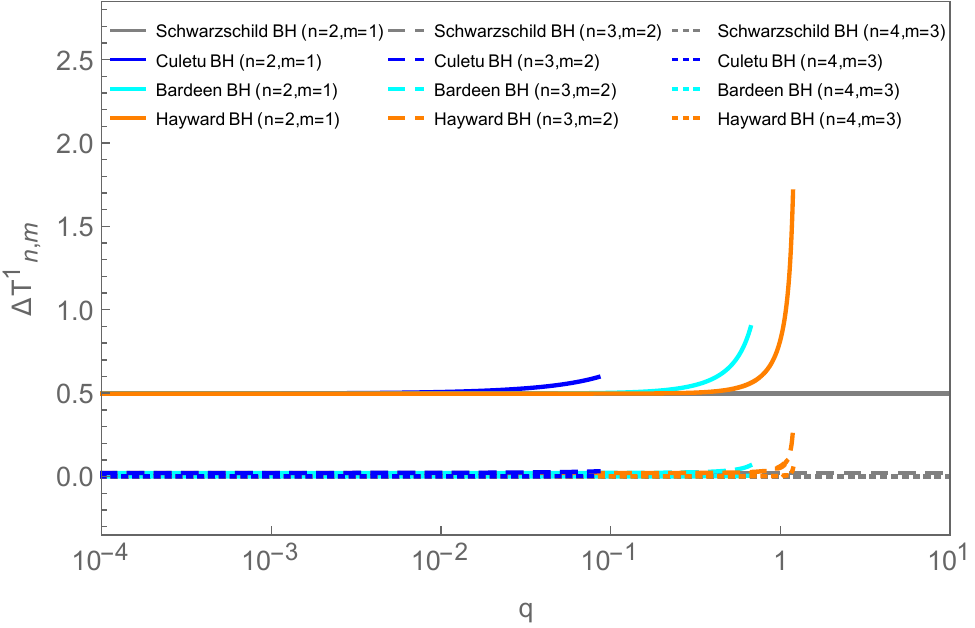}
	\includegraphics[width=8cm]{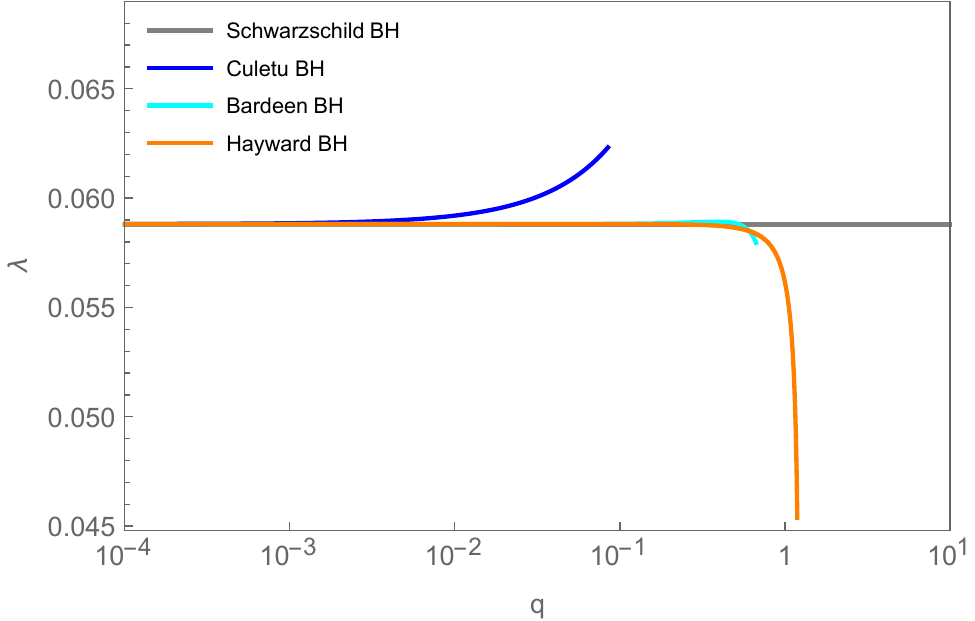}\\
	\caption{Same as Fig. \ref{rmb}, but for the angular separation $s$, magnification $r_{mag}$, subleading time delay term $\Delta T^1_{n,m}$ and QNM Lyapunov exponent $\lambda$.}
	\label{srd}
\end{figure*}
%%%%%%%%%%%%%%%%%%
%%%%%%%%%%%%%%%%%%%%%%%%%%%%%%%
\subsubsection{Constraining results by strong gravitational lensing}
To constrain the parameter $q$, we perform a $\chi^2$ analysis based on observables sensitive to the near horizon geometry \cite{Zhang:2023rsy}
\begin{eqnarray}
	\chi^2 = \sum_{n=1}^{2} \frac{\left(\theta_d^{\rm th} - \theta_d^{\rm obs}\right)^2}{\sigma^2}.
\end{eqnarray}
We adopt the EHT measurements of the angular shadow radius for M87* and Sgr A*, given by $\theta_d^{\rm obs} = 21 \pm 1.5\,\mu{\rm as}$ and $24.35 \pm 3.5\,\mu{\rm as}$, respectively.

The resulting constraints, summarized in Table~\ref{sigmas} and Fig.~\ref{chi2}, show that strong field shadow observations impose significantly tighter bounds on $q$ than those derived from the Einstein ring. This enhanced sensitivity arises because shadow measurements probe photon trajectories in the immediate vicinity of the event horizon.
Restricting to $q \geq 0$, we find that the allowed ranges shrink to $\mathcal{O}(10^{-2})$, $\mathcal{O}(10^{-1})$ and $\mathcal{O}(1)$ for the Culetu, Bardeen and Hayward black holes, respectively. More precisely, the $1\sigma$ and $2\sigma$ constraints on the parameter $q$ for the considered black hole  models are summarized as follows:
\begin{eqnarray}
	\label{strongconstraint}
	&&\text{Culetu BH}:\quad 0 \leq q < 0.0466 \ <0.0847 ; \nonumber \\
	&&\text{Bardeen BH}:\quad 0 \leq q < 0.5115 \ <0.6682  ; \nonumber \\
	&&\text{Hayward BH}: \quad 0 \leq q < 1.0258 \ <1.1881 .
\end{eqnarray}
The minimum of the $\chi^2$ function is located at $q = 0$ for all three models.
Among them, the Culetu black hole exhibits the most stringent constraint, indicating the highest sensitivity to $q$, while the Hayward model allows the largest parameter space.
%%%%%%%%%%%%%%%%%%%%%%%%%%%%%% 
\subsection{Joint lensing analysis}
We perform a joint constraint on $q$ using both Einstein ring and strong gravitational lensing. For the Culetu, Bardeen and Hayward black holes, the combined constraints are shown in Table~\ref{sigmas}. A comparison with the strong field only results shows they are essentially identical. This indicates that the posterior is dominated by strong field data, which implies that current galactic scale lensing observations lack sufficient precision to further tighten constraints on the regularization parameter.

%%%%%%%%%%%%%%%%%%%%%%%%%%%%%
\section{The model degeneracy}\label{degeneracy}
With the parameter space for the Culetu, Bardeen and Hayward black holes constrained by EHT observations (see Table~\ref{sigmas}), we proceed to characterize their respective strong field gravitational lensing features.

%%%%%%%%%%%%%%%%%%%%%
\subsection{Photon sphere radius $x_m$ and critical impact parameter $b_m$}
As illustrated in Fig.~\ref{rmb}, both the photon sphere radius $x_m$ and the critical impact parameter $b_m$ evolve as functions of the regularization parameter $q$, where they exhibit a monotonic decline and remaining consistently below their Schwarzschild counterparts ($q=0$).

For the photon sphere radius $x_m$, the three regular black hole models display distinct, model dependent trajectories. Although the curves follow a similar downward trend, they do not overlap in the region where $q$ increases toward the $2\sigma$ observational boundary ($q=q_{2\sigma}$), which leaves the $x_m$ values for the three models clearly separated. This divergence reflects the sensitive dependence of the photon sphere's location on the specific internal mathematical structure of each metric, particularly the $A'(x)/A(x)$ term.

Specially, the critical impact parameter $b_m$ exhibits a remarkable parallel evolution across the three models. Despite the underlying differences in their photon sphere radii, the $b_m$ curves follow nearly identical functional forms, which makes them appear virtually congruent throughout the considered parameter space. Notably, while $x_m$ values diverge at the $2\sigma$ limit, the corresponding $b_m$ values for all three black holes converge to almost the same numerical result at $q=q_{2\sigma}$ (see Table~\ref{tabtheta}).

This pronounced degeneracy in $b_m$ can be attributed to several factors. Firstly, the shared Schwarzschild asymptotics (see Sec.~\ref{BH}) ensure that the dominant gravitational contribution at these scales remains similar. Secondly, $b_m$ is a composite quantity involving both $x_m$ and the metric function $A(x_m)$. The model dependent variations in $x_m$ are effectively canceled out by corresponding changes in $A(x_m)$ through a compensation mechanism that suppresses inter model differences. Finally, current observational uncertainties further obscure these subtle deviations, which renders the $b_m$ dominated predictions effectively indistinguishable at the $2\sigma$ confidence level.
%%%%%%%%%%%%%%%%%%%%%
%%%%%%%%%%%%%%%%%%%%%%%%%%%

%%%%%%%%%%%%%%%%%%%%%%%%%%
\subsection{Strong lensing observables related to $b_m$}
Following Bozza's analytical formalism \cite{Bozza:2002zj}, we first evaluate the leading order strong lensing observables, which are essentially dictated by the critical impact parameter $b_m$. This shared dependence results in a pronounced geometric and temporal degeneracy, which renders different regular black hole candidates nearly indistinguishable under current observational precision.

This degeneracy is most evident in the shadow geometry. As illustrated in Table~\ref{tabtheta} and Fig.~\ref{theta}, both the angular radius $\theta_d$ and the shadow area $\tilde{A}$ exhibit a monotonic decline with increasing regularization parameter $q$. While M87* displays a smaller angular radius than Sgr A* due to its greater luminosity distance $D_L$, the trajectories for the Culetu, Bardeen and Hayward metrics follow nearly identical functional forms. This parallel evolution, combined with significant numerical overlap, makes these models indistinguishable within current $2\sigma$ uncertainties. A similar trend persists for the shadow area $\tilde{A}$, with its quadratic dependence on $b_m$ leading to a more rapid decline than that of $\theta_d$ while maintaining the structural resemblance across models. Such behavior suggests that current geometric measurements primarily probe the ``effective gravitational cross section'' near the photon sphere, rather than the microscopic mathematical details of the underlying core corrections.

The resulting ``macroscopic universality'' is further confirmed by the shadow cast presented in Fig.~\ref{cast}. Within $1\sigma$ and $2\sigma$ confidence intervals, the shadow outlines contract as $q$ increases, which results in a maximum in the Schwarzschild limit ($q=0$) and a minimum at the $2\sigma$ boundary ($q=q_{2\sigma}$). Because shadow morphology is governed by $R_{sh} \equiv b_m$ (see Eq.~(\ref{castEq})), and $b_m$ itself remains highly degenerate within existing constraints, the geometric profile alone is insufficient to resolve disparities in the underlying spacetime structures.

Crucially, this model degeneracy extends beyond static geometry into temporal and dynamical  features. As shown in Table~\ref{tabtheta} and Fig.~\ref{theta}, the leading order time delay $\Delta T^0_{n,m}$ and the real part of the quasinormal mode (QNM) frequency $\Omega_m$ are both analytically linked to $b_m$ by
\begin{eqnarray}
	&&\Delta T^0_{n,m} = 2\pi (n-m) b_m,\\
	&&\Omega_m = \frac{1}{b_m}.
\end{eqnarray}
For the most observationally relevant image orders, $(n,m) = (2,1), (3,2), (4,3)$, the time delay simplifies to $2\pi b_m$ and decreases with $q$, while $\Omega_m$ conversely increases. Despite their opposing slopes, both observables exhibit the same parallel evolution across the three regular black hole models. This congruence indicates that incorporating basic time delay or QNM data at current precision levels cannot break the degeneracy, as these dynamical signatures essentially probe the same physical degree of freedom as the shadow radius.
%%%%%%%%%%%%%%%%%%%%%%%%%%
\subsection{Summary of existing constraints}
In summary, while the photon sphere radius $x_m$ is sensitive to metric derivatives, the critical impact parameter $b_m$ is a composite quantity where model dependent variations are largely canceled out. This interplay, amplified by current experimental uncertainties, leads to the observed degeneracy in shadow size, area and fundamental frequencies. To distinguish these models, it is essential to introduce independent observables sensitive to higher order structures or dynamical properties, as discussed in the following section.
%%%%%%%%%%%%%%%%%%%%%%%%%%%%%%%%%%%%%
%%%%%%%%%%%%
\begin{figure*}[!htb]
	\includegraphics[width=8cm]{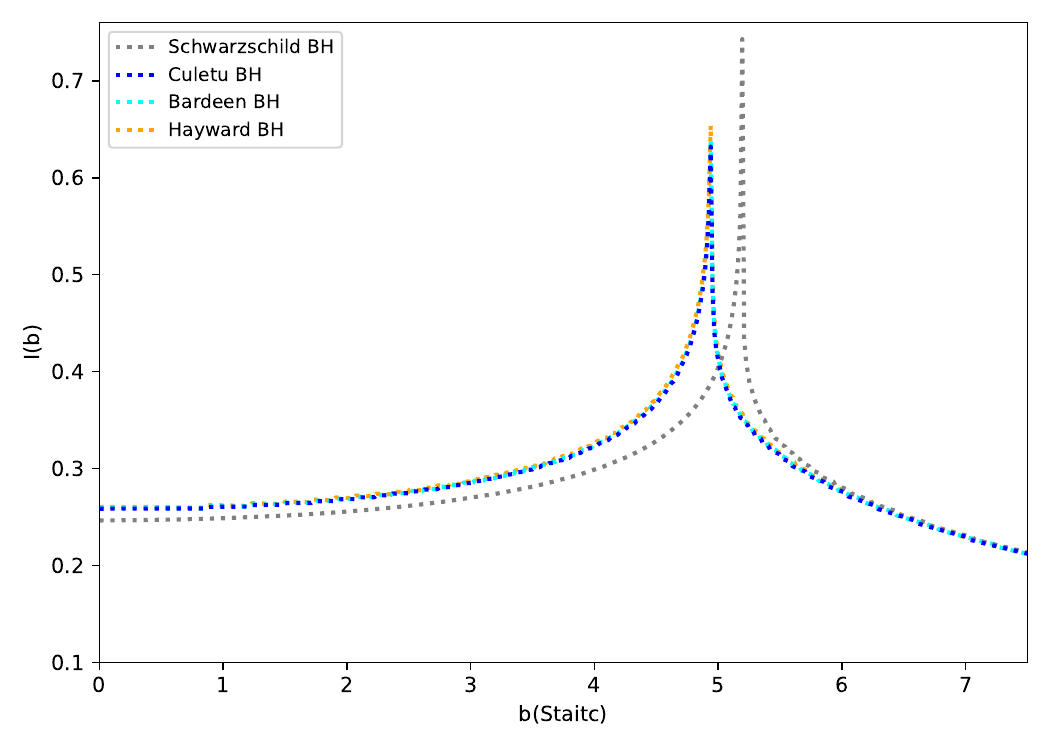}
	\includegraphics[width=8cm]{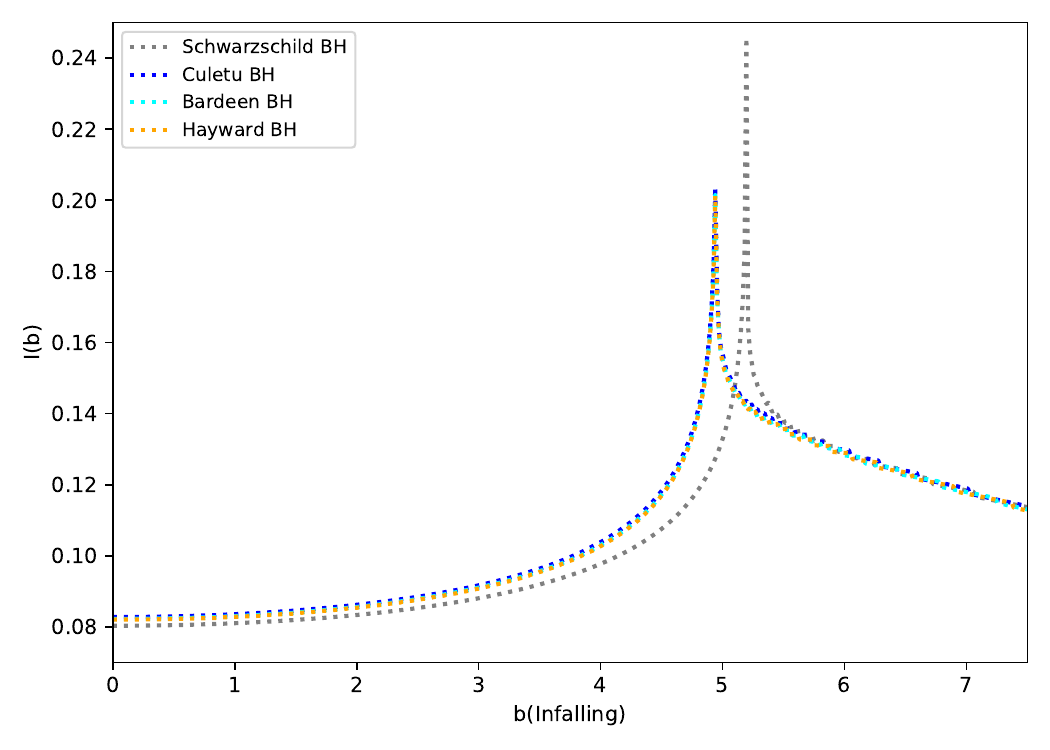}
	\caption{Specific intensity profiles $I(b)$ for Culetu, Bardeen and Hayward black holes at the $1\sigma$ constraint $q=q_{1\sigma}$. The left panel illustrates static accretion flow case, while the right panel shows infalling accretion case. The Schwarzschild result is included as standard reference for comparison.}
	\label{intensity}
\end{figure*}
%%%%%%%%%%%%
%%%%%%%%%%%%%%%
\begin{figure*}[!htb]
	\includegraphics[width=7cm]{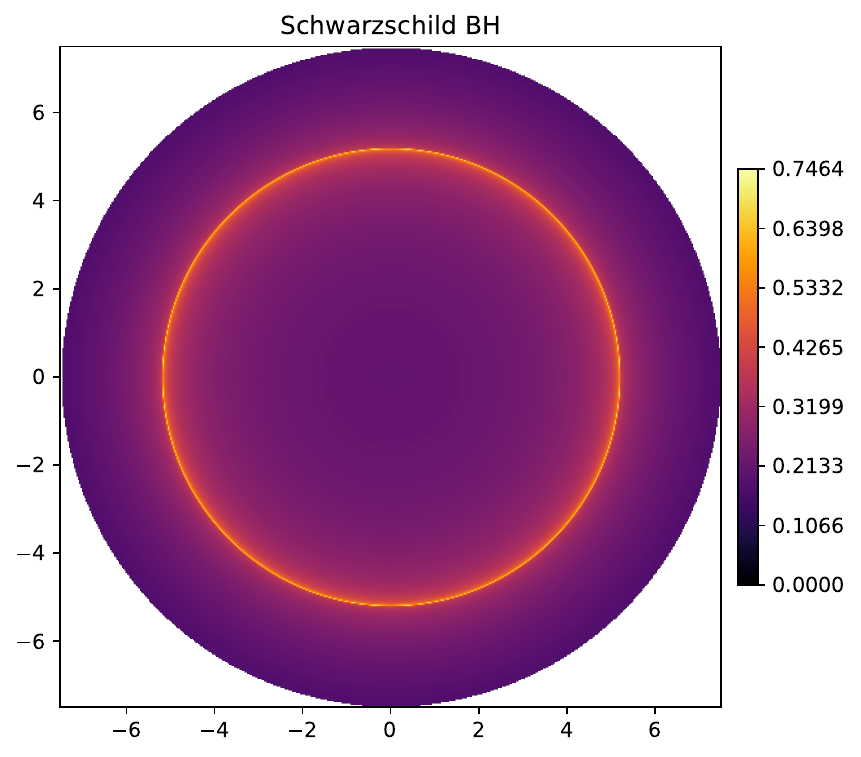}
	\includegraphics[width=7cm]{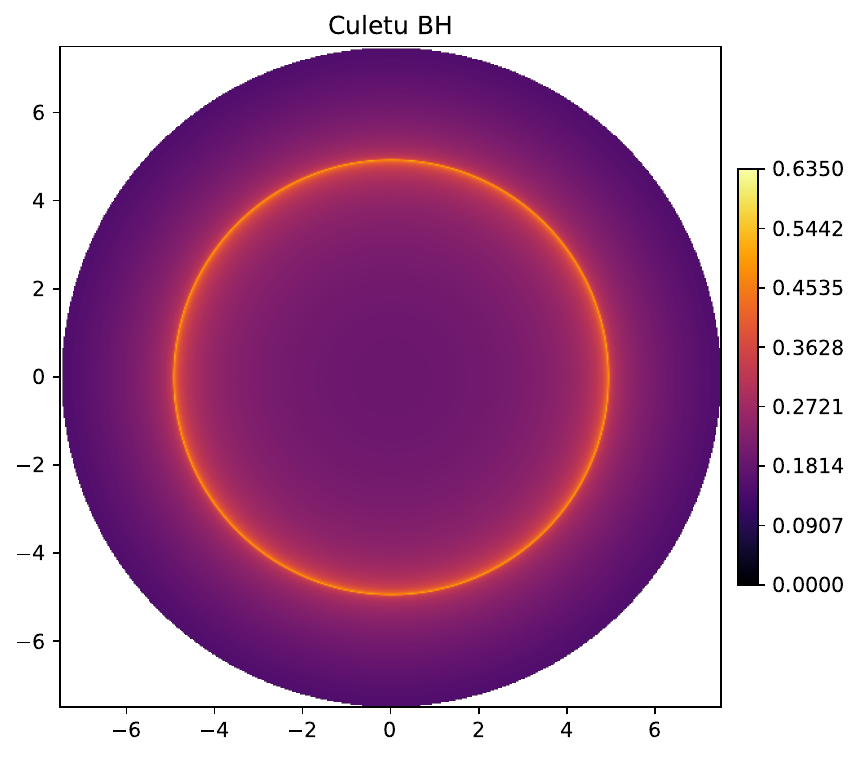}\\
	\includegraphics[width=7cm]{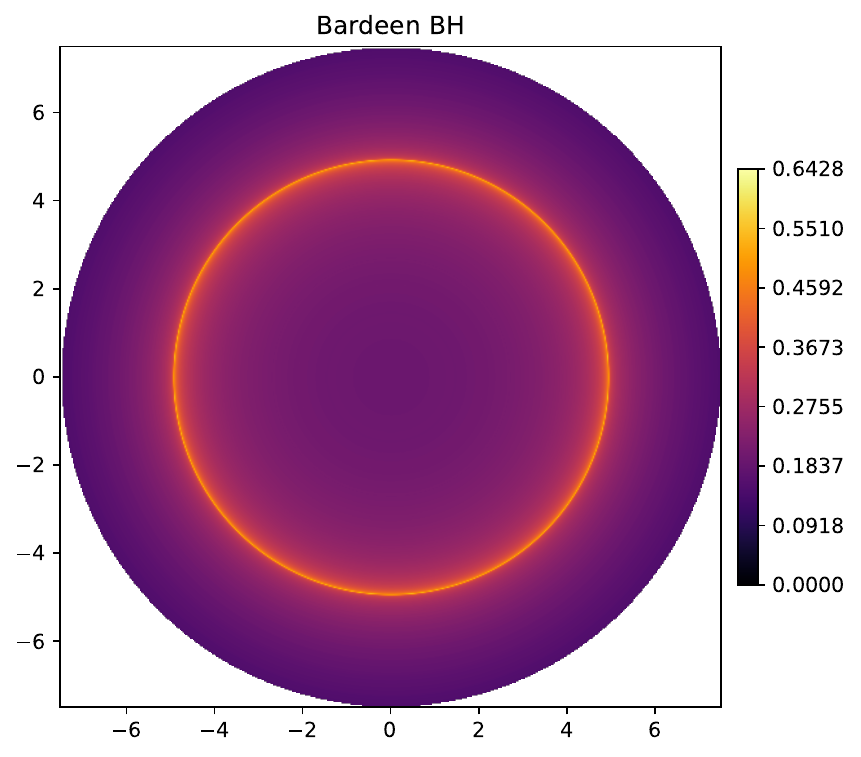}
	\includegraphics[width=7cm]{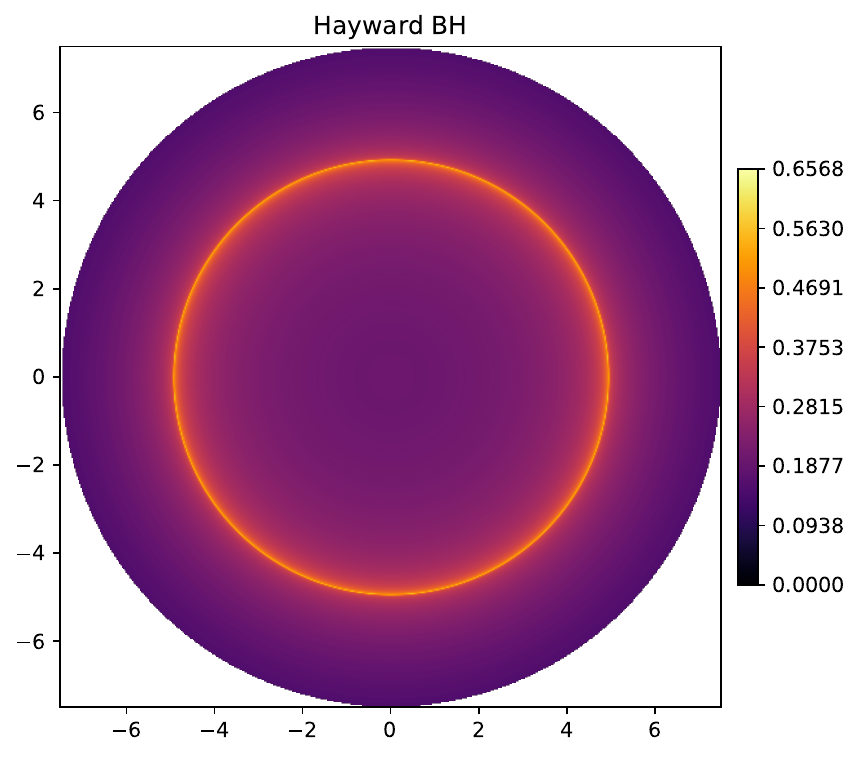}
	\caption{Similar as the left panel of Fig~\ref{intensity}, but showing the two dimensional images of shadow and photon ring for Culetu, Bardeen and Hayward models. Note that the colorbars are normalized individually, with the maximum value in each scale corresponding to the peak specific intensity of respective black hole model.}
	\label{static}
\end{figure*}
%%%%%%%%%%%%
%%%%%%%%%%%%%%%
\begin{figure*}[!htb]     
	\includegraphics[width=7cm]{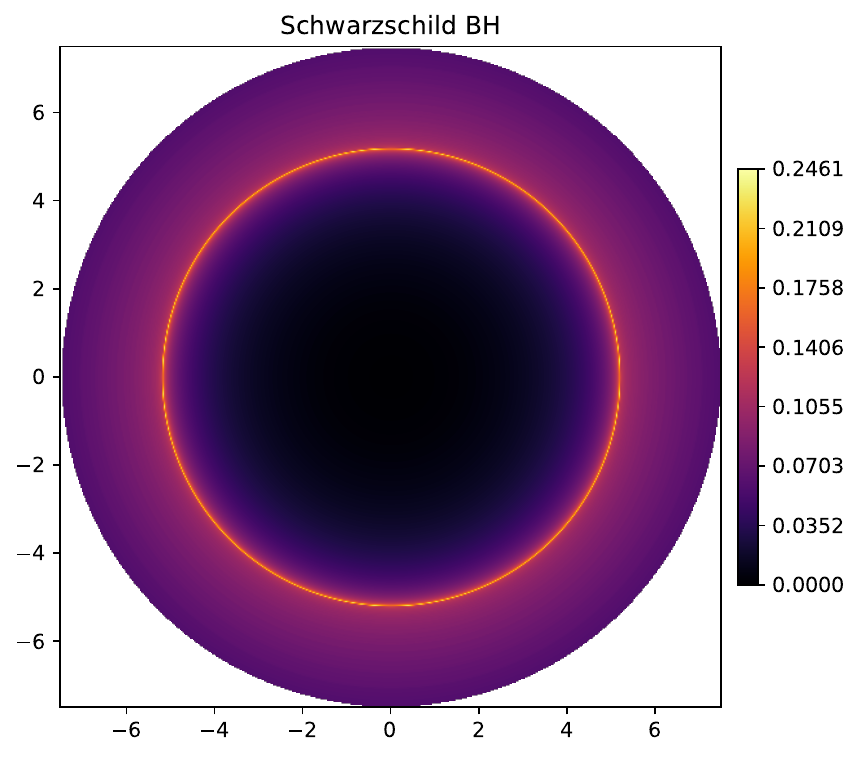}
	\includegraphics[width=7cm]{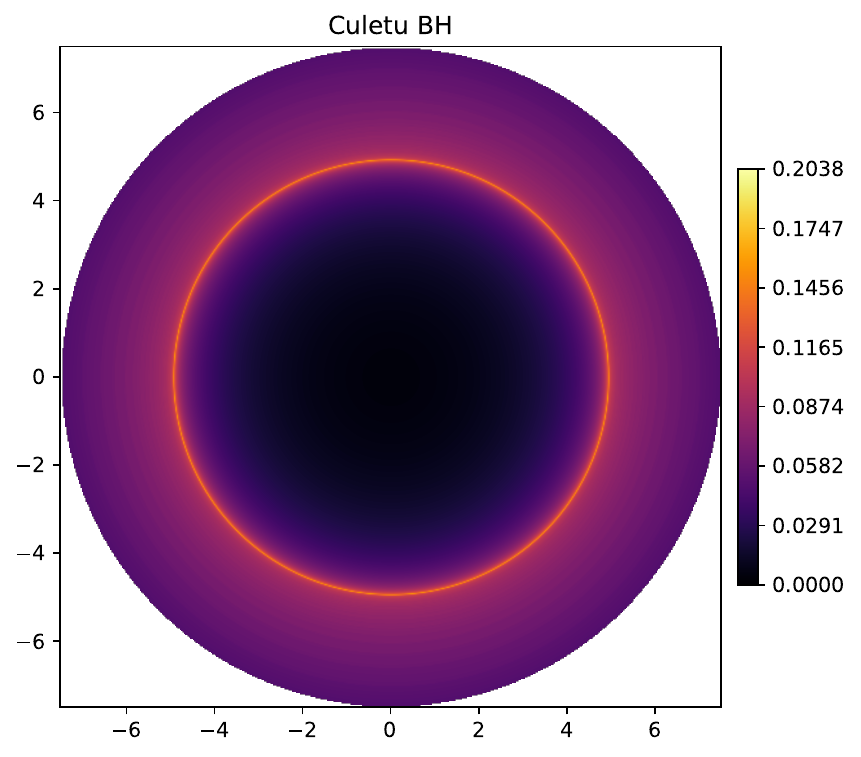}\\
	\includegraphics[width=7cm]{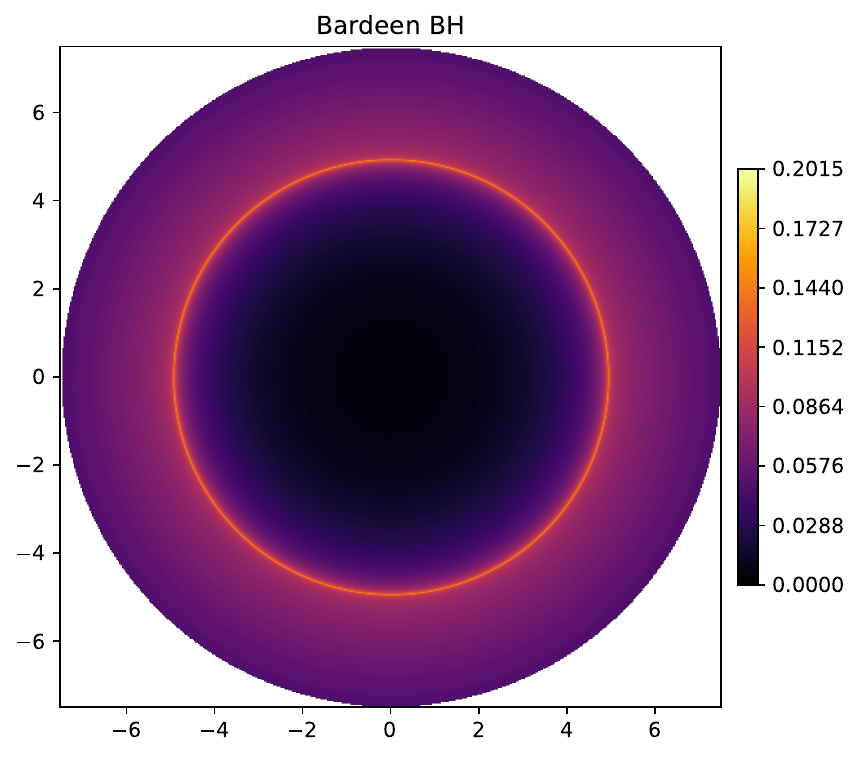}
	\includegraphics[width=7cm]{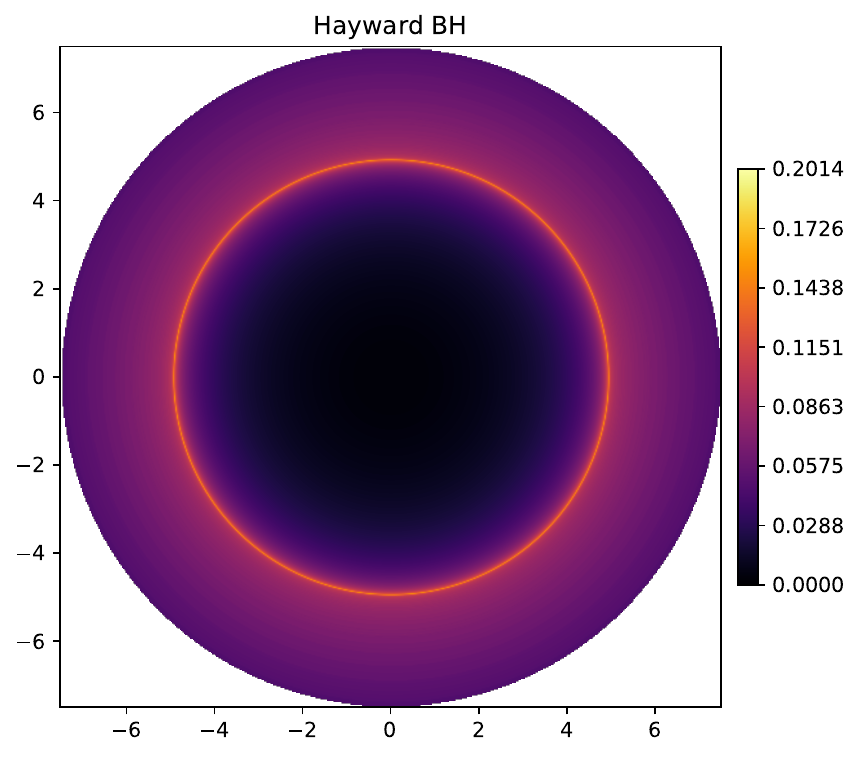}
	\caption{Same as Fig. \ref{static}, but for the infalling accretion flow. }
	\label{infalling}
\end{figure*}
%%%%%%%%%%%%%%%%%%%%%%%%%%
%%%%%%%%%%%%%%%%%%%%%%%%%%%%%%%%%%%%%%
\section{Potential avenues to break model degeneracy}\label{observations}
To move beyond the limitations of $b_m$ dominated observables, we explore higher order signatures that probe the fine grained curvature structure near the photon sphere which may be resolvable with future high precision observations.
%%%%%%%%%%%%%%%%%%%%%%%%%%%%%%
\subsection{Subleading observables}
To resolve the microscopic disparities in regular black hole spacetimes, we move beyond the leading order impact parameter to examine the strong lensing coefficients $\bar{a}$ and $\bar{b}$, which characterize the logarithmic divergence of the deflection angle near the photon sphere (see ~\ref{appBozza}). As shown in Fig.~\ref{a}, the coefficient $\bar{a}$ that reflects photon orbit instability exhibits a monotonic increase with $q$. While the Bardeen and Hayward models follow nearly identical trajectories, the Culetu model shows a distinct deviation, which suggests a unique internal scaling. This divergence is even more pronounced in the coefficient $\bar{b}$, which exhibits a bifurcated behavior where it increases above the Schwarzschild baseline for the Culetu model and decreases for the Bardeen and Hayward cases. Since $\bar{b}$ encodes global information integrated along the null trajectory, these distinct signatures offer a robust mechanism to break the macroscopic degeneracies observed in the previous section.

These coefficients directly govern the primary observational features of the relativistic images, which start with the angular separation $s$ between the first relativistic image and the asymptotic position $\theta_{\infty}$
\begin{eqnarray}
	s = \theta_1 - \theta_{\infty} = \theta_{\infty} \exp\left(\frac{\bar{b}}{\bar{a}} - \frac{2\pi}{\bar{a}}\right).
\end{eqnarray}
As illustrated in Fig.~\ref{srd}, $s$ increases monotonically with the regularization parameter $q$. The quantitative discrepancies emerging between the three regular black hole models indicate that $s$ is highly sensitive to the ratio $\bar{b}/\bar{a}$, which makes it an effective probe of the detailed geometry near the critical orbit. Complementing this positional data is the magnification ratio $r_{\rm mag}$, which characterizes the brightness contrast between the first image and the sum of all higher order images
\begin{equation}
	r_{\rm mag} = 2.5 \log_{10} \left(\frac{\mu_1}{\sum_{n=2}^{\infty}\mu_n}\right) = 2.5 \log_{10} \left[ \exp\left(\frac{2\pi}{\bar{a}}\right) \right].
\end{equation}
Unlike the angular separation $s$, $r_{\rm mag}$ depends solely on the coefficient $\bar{a}$ and decreases as $q$ increases. By isolating the impact of orbit instability, $r_{\rm mag}$ provides independent constraints on the spacetime structure. However, resolving these features sets a high bar for future facilities such as the next generation Event Horizon Telescope. This goal requires an angular resolution below $1\,\mu\mathrm{as}$ and photometric precision better than $0.1\,\mathrm{mag}$.

Further dynamical insights are provided by the subleading correction to the time delay $\Delta T^1_{n,m}$, which is defined as
\begin{eqnarray}
	\Delta T^1_{n,m} = 2\sqrt{\frac{B_m}{A_m}}\sqrt{\frac{b_m}{\tilde{c}}} e^{\frac{\bar{b}}{2\bar{a}}}\left( e^{-\frac{m\pi}{\bar{a}}} - e^{-\frac{n\pi}{\bar{a}}}\right).
\end{eqnarray}
Further dynamical insights are provided by the subleading correction to the time delay $\Delta T^1_{n,m}$, which involves a non-trivial combination of metric functions and their second derivatives. Adopting geometric units $GM_{\bullet}/c^3$, we observe in Fig.~\ref{srd} that $\Delta T^1_{n,m}$ decreases monotonically with $q$ and exhibits a strong dependence on the image order $(n,m)$. Specifically, at the $1\sigma$ confidence level, the magnitudes of these corrections scale as $\mathcal{O}(1)$ for $\Delta T^1_{2,1}$, $\mathcal{O}(10^{-1})$ for $\Delta T^1_{3,2}$ and $\mathcal{O}(10^{-2})$ for $\Delta T^1_{4,3}$. This rapid suppression across successive orders, coupled with a high sensitivity to near horizon curvature, establishes a clear hierarchy wherein the Hayward model yields the largest correction, followed by the Bardeen and Culetu models, respectively. Such a progression captures fine grained geometric features that the leading order impact parameter $b_m$ fails to resolve.

Such orbital dynamics are intrinsically linked to the Lyapunov exponent in the eikonal limit
\begin{eqnarray}
	\lambda = \frac{1}{\bar{a} b_m}.
\end{eqnarray}
Physically, $\lambda$ corresponds to the imaginary part of the quasinormal mode frequency and measures the divergence rate of nearby null geodesics. As illustrated in Fig.~\ref{srd}, $\lambda$ increases with $q$ in the Culetu model which indicates enhanced instability. In contrast, it decreases in the Hayward model which suggests relative stabilization. The Bardeen model, meanwhile, exhibits non-monotonic behavior that reflects the physical competition between a steepening effective potential and the emergence of a diffuse de Sitter-like core. 

By encoding these high order geometric and dynamical details, the suite of observables $\{s, r_{\rm mag}, \Delta T^1_{n,m}, \lambda\}$ significantly expands the parameter space available for testing gravity and distinguishing between competing regular black hole models.
%%%%%%%%%%%%%%%%%%%%%%%%%%%%%%%%%%%%%

%%%%%%%%%%%%%%%%%%%%%%%%%%%%%%%%%%%%%
\subsection{Shadows with spherical accretion flows}
We investigate the shadows of regular black holes immersed in both static and infalling spherical accretion flows. Restricted to the equatorial plane ($\theta = \pi/2$), the observed specific intensity is determined by integrating the emissivity along photon trajectories, which accounts for gravitational redshift \cite{Jaroszynski:1997bw}
\begin{eqnarray}
	I(\nu_{\rm obs}) = \int g^3 j(\nu_e) dl_{\rm prop},
\end{eqnarray}
where $j(\nu_e)$ is the specific emissivity in the rest frame, $dl_{\rm prop}$ is the infinitesimal proper length, and the redshift factor is defined by the four-velocity of the plasma $u_e^\mu$ and the observer $u_{obs}^\mu = (1, 0, 0, 0 )$ as
\begin{eqnarray}
	\label{gfactor}
	g = \frac{k_\mu u^\mu_{\text{obs}}}{k_\nu u^\nu_e}.
\end{eqnarray}

In a static flow, the plasma four-velocity relative to a static observer is $u_e^\mu = \left(1/\sqrt{A(x)}, 0, 0, 0 \right)$, which simplifies the redshift factor to $g = \sqrt{A(x)}$. Assuming a monochromatic emissivity $j(\nu_e) \propto \delta(\nu_e - \nu_s) / x^2$, the observed intensity becomes
\begin{eqnarray}
	I (b) = \int \frac{A(x)^{3/2}}{x^2} \sqrt{\frac{1}{A(x)} + \frac{b^2}{x^2 - b^2 A(x)}} , dx.
\end{eqnarray}
As shown in Fig.~\ref{intensity} (left), $I(b)$ peaks sharply at the photon ring $b = b_m$. Despite the geometric degeneracy where the shadow radii $R_{sh}$ for Culetu, Bardeen and Hayward models are virtually indistinguishable, a clear hierarchy emerges in the specific intensity: $I_{\rm Schwarzschild} > I_{\rm Hayward} > I_{\rm Bardeen} > I_{\rm Culetu}$. In this lensing dominated regime, the Schwarzschild spacetime's superior lensing allows photons to traverse longer, more curved paths through the high emissivity region, while the Culetu model's weaker lensing leads to more direct trajectories and lower accumulated intensity.

This ordering undergoes a fundamental inversion when considering an infalling accretion flow, where the plasma four-velocity is $u_e^\mu =(1/A(x),-\sqrt{1-A(x)},0,0)$. The redshift factor and resulting intensity are given by
\begin{eqnarray}
	I (b) = \int \frac{g^3}{x^2 \sqrt{\frac{1}{A(x)} \left(\frac{1}{A(x)}-\frac{b^2}{x^2}\right)}}  dx,
\end{eqnarray}
where
\begin{eqnarray}
	g = \frac{A(x)}{1+\sqrt{(1-\frac{b^2}{x^2}A(x))(1-A(x))}}.
\end{eqnarray}
As illustrated in Fig.~\ref{intensity} (right) and Fig.~\ref{infalling}, the infalling motion shifts the system into a dynamics dominated regime. The rapid radial plunge toward the horizon induces severe Doppler redshift, which reduces peak intensities by a factor of three and darkening the shadow interior. Crucially, the brightness hierarchy reorders to $I_{\rm Schwarzschild} > I_{\rm Culetu} > I_{\rm Bardeen} > I_{\rm Hayward}$. In this regime, the Culetu model surpasses its regular black hole counterparts because its weaker lensing facilitates a more favorable alignment between photon paths and the infalling velocity field, which results in less severe Doppler suppression. This transition from lensing induced amplification to motion induced dimming effectively breaks the geometric degeneracy of the shadow, which provides a robust observational channel to distinguish between regular black hole spacetimes and aligning more closely with the high contrast features observed by the EHT for M87* and Sgr A*.
%%%%%%%%%%%%%%%%%%%%%%%%%%%%%%%%%%%%%
\subsection{Probing regular black holes with ngEHT}
The high order signatures $\{s, r_{\rm mag}, \Delta T^1_{n,m}, \lambda\}$ established in this work provide a concrete benchmark for the ngEHT. While $b_m$ exhibits ``macroscopic universality,'' these sub-leading corrections are sensitive to the fine grained curvature of non-singular cores. With $\sim 15\,\mu\mathrm{as}$ resolution and $1000:1$ dynamic range \cite{Johnson:2023ynn}, the ngEHT can resolve $n=1, 2$ lensing sub-rings, allowing a direct measurement of the Lyapunov exponent $\lambda$ to distinguish between the unstable orbits of Culetu and the relatively stabilized Hayward geometry. Furthermore, observations at $345 GHz$ will exploit a more transparent plasma regime \cite{Johnson:2023ynn}, bypassing the optically thick layers that obscure near horizon physics. This transparency, combined with high cadence monitoring, enables the detection of the predicted brightness hierarchy inversion in ``black hole movies.'' Since the Culetu model's unique lensing velocity alignment results in less severe Doppler suppression, its relative intensity provides a robust channel to break geometric degeneracies and distinguish between specific regular black hole formulations.
%%%%%%%%%%%%%%%%%%%%%%%%%%%%%%%%%%%%%
\section{Conclusion}\label{conclusion}
This work systematically examined the observational signatures of Culetu, Bardeen and Hayward regular black holes across lensing, shadow, QNM and accretion regimes. Our findings reveal a significant disparity in constraining power between different observational scales. While the Einstein ring data alone provide only extremely loose constraints, with the regularization parameter $q$ exceeding $\mathcal{O}(10^3)$. The strong field EHT data yield much more stringent $2\sigma$ limits: $0 \leq q < 0.0466 <0.0847$ for Culetu, $0 \leq q < 0.5115 <0.6682$ for Bardeen and $0 \leq q < 1.0258 <1.1881$ for Hayward. Notably, a joint analysis combining the Einstein ring and strong lensing observables yields constraints consistent with those derived from the strong field regime alone, which confirms that the latter entirely dominates the available parameter space and that the inclusion of Einstein ring data does not improve the overall precision.

Despite these bounds, leading order geometric observables, including shadow radius $b_m(R_{sh})$, angular size $\theta_{\infty}(\theta_d)$ and the quasinormal mode frequency $\Omega_m$, remain highly degenerate within the allowed $2\sigma$ range. This ``macroscopic universality'' indicates that current observations primarily probe the effective gravitational cross section rather than the specific mathematical details of non-singular core corrections.

To break this degeneracy, we identify high order signatures such as the Lyapunov exponent $\lambda$, angular separation $s$ and subleading time delay $\Delta T^1_{n,m}$ as sensitive probes of near horizon curvature. Crucially, we demonstrate that the brightness hierarchy of accretion induced intensity profiles $I(b)$ undergoes a fundamental inversion when transitioning from lensing dominated static flows to dynamics dominated infalling flows. These subtle disparities set a concrete target for future high resolution instruments like the next generation Event Horizon Telescope to distinguish between competing regular black hole models.
%%%%%%%%%%%%%%%%%%%%% 
\section{Acknowledgements}
The authors thanks Hui Li, Dongze He and Xuanjun Niu for the valuable comments and suggestions that helped improve this manuscript. The authors are supported by National Natural Science Foundation of China under Grant No.12275037 and No.12275106.
%%%%%%%%%%%%%%%%%%%%%%%
\appendix
\section{WEAK FIELD DEFLECTION ANGLE AND EINSTEIN RING}\label{appweak}
In this section, we briefly present the derivation of weak deflection angle $\alpha(b)$ and subsequent determination of Einstein radius $\theta_E$ for black hole metrics considered in this work. Following the methodology of Ref.~\cite{Keeton:2005jd}, we perform a post-Newtonian expansion in the weak field limit ($x_0 \gg 1$) to express the deflection angle as
\begin{align}
		\alpha(x_0) = &\frac{4}{x_0} + \frac{15\pi/4 - 4 + A_2}{x_0^2} + \frac{122/3 - 15\pi/2 + A_3}{x_0^3} \nonumber \\
		&+ \frac{-130 + 3465\pi/64 + A_4}{x_0^4} + \mathcal{O}(x_0^{-5}),
\end{align}
The model specific coefficients are given by
\begin{itemize}
	[label=\textbullet]
	\item Culetu BH: $A_2 = -9q\pi/2$, $A_3 = 9q\pi - 84q + 32q^2$ and $A_4 = -2475q\pi /16 + 1773q^2\pi/16 - 75q^3\pi/4 + 300q - 120q^2$;
	\item Bardeen BH: $A_2 = 0$, $A_3 = -8q^2$ and $A_4 = -315q^2\pi/16 + 30q^2$;
	\item Hayward BH: $A_2 = 0$, $A_3 = 0$ and $A_4 = -15q^3\pi/8$.
\end{itemize}
Expressing the deflection angle in terms of impact parameter $b$, we obtain
\begin{align}
		\label{weak}\alpha(b) = &\frac{4M}{b} + \left(\frac{15\pi}{4} + B_2\right) \frac{M^2}{b^2} + \left(\frac{128}{3}+B_3\right) \frac{M^3}{b^3} \nonumber \\
		&+ \left(\frac{3465\pi}{64}+B_4\right) \frac{M^4}{b^4} + \mathcal{O}(b^{-5}),
\end{align}
where the coefficients are
\begin{itemize}
	[label=\textbullet]
	\item Culetu BH: $B_2 = - 9q\pi/2$, $B_3 = -96q+32q^2$ and $B_4 = -2835q\pi/16+2205q^2\pi/16+75q^3\pi/4$;
	\item Bardeen BH: $B_2 = 0$, $B_3 = -8q^2$ and $B_4 = 315q^2\pi/16$;
	\item Hayward BH: $B_2 = 0$, $B_3 = 0$ and $B_4 = -15q^3\pi/8$.
\end{itemize}
For a perfectly aligned source--lens--observer configuration, the Einstein ring angle $\theta_E$ satisfies the lens equation
\begin{eqnarray}
	\theta_E = \frac{d_{LS}}{d_S} \alpha(b).
	\label{eq:lens_E}
\end{eqnarray}
In the following, we obtain $\theta_E$ by solving Eq.~(\ref{eq:lens_E}) perturbatively. Using weak field expansion of deflection angle and relation $b=d_L\theta_E$, Eq.~\eqref{eq:lens_E} can be rewritten as
\begin{eqnarray}
	\theta_E^5 = \delta \theta_E^3 + \gamma \theta_E^2 + \mu \theta_E + \nu,
	\label{eq:theta_cubic_eps}
\end{eqnarray}
where $\delta = 4Md_{LS}/(d_L d_S)$, $\gamma = \eta M^2 d_{LS}/(d_L^2 d_S)$ with $\eta=15\pi/4 + B_2$, $\mu = B_3 M^3 d_{LS}/(d_L^3 d_S)$ and $\nu = B_4 M^4 d_{LS}/(d_L^4 d_S)$. To solve Eq.~\eqref{eq:theta_cubic_eps} perturbatively, we expand the Einstein ring angle as
\begin{eqnarray}
	\theta_E = \theta_{E1} + \theta_{E2} + \theta_{E3} + \theta_{E4} + \mathcal{O}(\epsilon^5).
	\label{eq:theta_expand_eps}
\end{eqnarray}
At the leading order of the expansion, Eq.~\eqref{eq:theta_cubic_eps} reduces to the relation $\theta_{E1}^5 = \delta \theta_{E1}^3$, from which the first order term is obtained
\begin{eqnarray}
	\theta_{E1} = \sqrt{\frac{4M d_{LS}}{d_L d_S}}.
	\label{eq:theta1_eps}
\end{eqnarray}
Substituting the expansion back into the governing equation and collecting terms of the next relevant order, $2\theta_{E1}^4 \theta_{E2}=\gamma \theta_{E1}^2$, we obtain the next order contribution
\begin{eqnarray}
	\theta_{E2} = \frac{\gamma}{2\delta}.
	\label{eq:theta2_eps}
\end{eqnarray}
Continuing this procedure, the subsequent term is governed by the relation $2\theta_{E1}^2\theta_{E3} = 3 \theta_{E1}^2 \theta_{E2}^2 + \mu - 6 \theta_{E1}^3 \theta_{E2}^2$, from which the corresponding contribution follows
\begin{eqnarray}
	\theta_{E3} = \frac{3 \gamma^2}{8\delta^2} + \frac{\mu}{2\delta} - \frac{3\theta_{E1} \gamma^2}{4\delta^2}.
	\label{eq:theta3_eps}
\end{eqnarray}
Following the same iterative procedure, the final contribution can be obtained as
\begin{align}
		\theta_{E4} = &\frac{1}{16 \delta^{9/2}} \Big[ -6\delta^2\gamma^2 - 6\delta^{5/2}\gamma^2 + 12\delta^3\gamma^2 + 9\gamma^3 \nonumber \\
		&- 25\delta^{1/2}\gamma^3 - 24\delta\gamma^3 + 48\delta^{3/2}\gamma^3 + 8\delta^3\mu
		- 8\delta^{7/2}\mu \nonumber \\
		&+ 12\delta \gamma \mu + 4\delta^{3/2}\gamma\mu 
		- 32\delta\gamma\mu + 8\delta^{5/2}\nu \Big].
	\label{eq:theta4_eps}
\end{align}
Combining the expansion in Eq.~(\ref{eq:theta_expand_eps}) with the specific corrections given in Eqs.~(\ref{eq:theta1_eps})--(\ref{eq:theta4_eps}), one can determine the numerical value of the Einstein ring angle $\theta_E$ for Culetu, Bardeen and Hayward regular black holes.
%%%%%%%%%%%%%%%%%%%%%%%%%%%%%%%
\section{THE BOZZA'S PROCEDURE}\label{appBozza}
In this appendix we briefly review Bozza's analytical formalism for strong gravitational lensing \cite{Bozza:2002zj}. A photon orbiting a black hole has an innermost unstable circular orbit known as the photon sphere, with radius $x_m$. The impact parameter $b$ is conserved along the trajectory and relates to the closest approach $x_0$ as $b = \sqrt{C_0/A_0}$, where the subscript ``$0$'' denotes evaluation at $x_0$. The critical impact parameter corresponds to the photon sphere and is given by $b_m = \sqrt{C_m/A_m}$, where ``$m$'' indicates evaluation at $x_m$.
Introducing the variable $y = A(x)$ and $z = (y - y_0)/(1 - y_0)$, the deflection integral can be rewritten as
\begin{eqnarray}
	\tilde{I}(x_0) = \int_0^1 R(z,x_0) f(z,x_0) dz,
\end{eqnarray}
where
\begin{eqnarray}
		R(z,x_0) = \frac{2 \sqrt{B y}}{C A'} (1-y_0)\sqrt{C_0},
\end{eqnarray}
\begin{eqnarray}
		f(z,x_0) = \frac{1}{\sqrt{y_0 - [(1-y_0)z + y_0] C_0/C}}
\end{eqnarray}
where $R(z,x_0)$ is regular, while $f(z,x_0)$ diverges as $z \to 0$. Expanding the square root of $f(z,x_0)$ to second order yields
\begin{eqnarray}
	f(z,x_0) \sim f_0(z,x_0) = \frac{1}{\sqrt{\tilde{\alpha} z + \tilde{\beta} z^2}},
\end{eqnarray}
with
\begin{eqnarray}
	\begin{aligned}
		\tilde{\alpha} = \frac{1-y_0}{C_0 A'_0}\left(C'_0 y_0 - C_0 A'_0\right), 
	\end{aligned}
\end{eqnarray}
\begin{align}
		\tilde{\beta} = &\frac{(1-y_0)^2}{2 C_0^2 {A'_0}^3} \Big[ 2 C_0 C'_0 {A'_0}^2 + (C_0 C''_0 - 2 {C'_0}^2)y_0 A'_0 - \nonumber \\
		&\quad C_0 C'_0 y_0 A''_0 \Big].
\end{align}
The integral is split into a divergent and a regular part
\begin{eqnarray}
	\tilde{I}(x_0) = \tilde{I}_D(x_0) + \tilde{I}_R(x_0),
\end{eqnarray}
with
\begin{eqnarray}
		\tilde{I}_D(x_0) = \int_0^1 R(0,x_m) f_0(z,x_0) dz, 
\end{eqnarray}
\begin{equation}
		\tilde{I}_R(x_0) = \int_0^1 \big(R(z,x_m) f(z,x_m) - R(0,x_m) f_0(z,x_m)\big) dz.
\end{equation}
Expanding the impact parameter near the photon sphere gives
\begin{eqnarray}
	b - b_m = \tilde{c} (x_0 - x_m)^2,
\end{eqnarray}
where $\tilde{c} = (C''_m y_m - C_m A''_m)/4 \sqrt{y_m^3 C_m}$.
Finally, the deflection angle in the strong field limit can be expressed as
\begin{eqnarray}
	\alpha(\theta) = -\bar{a} \ln \left( \frac{b}{b_m} - 1 \right) + \bar{b},
\end{eqnarray}
with
\begin{eqnarray}
		\bar{a} = \frac{R(0,x_m)}{2 \sqrt{\beta_m}}, 
\end{eqnarray}
\begin{eqnarray}
		\bar{b} = -\pi + \bar{a} \ln \frac{2 \beta_m}{y_m} + b_R, 
\end{eqnarray}
where $b_R = \tilde{I}_R(x_m)$, $\beta_m = 2 \tilde{c} C_m^{3/2} y_m^{-1/2} (1 - y_m)^2$.
And, $\bar{a}$ and the first term of $\bar{b}$ encode the divergence, while $b_R$ represents the regular part. 
%%%%%%%%%%%%%%%%%%%%%%%%%%%%%%%

\bibliographystyle{apsrev4-2}
\bibliography{regularBH0326}% Produces the bibliography via BibTeX.

\end{document}